\documentclass[pra,final,floatfix,twocolumn,showpacs]{revtex4-1}

\usepackage{graphicx}
\usepackage{dcolumn}
\usepackage{bm}
\usepackage{amssymb}
\usepackage{natbib}
\usepackage{amsmath}
\usepackage{color}
\usepackage{datetime}
\usepackage{footnote}
\usepackage{bbold}
\usepackage[normalem]{ulem}
\usepackage{dcolumn}
\usepackage{ragged2e}

\newcommand{\ket}[1]{\ensuremath{|#1\rangle}}
\newcommand{\bra}[1]{\ensuremath{\langle#1|}}

\newcommand{\be}{\begin{equation}}
\newcommand{\ee}{\end{equation}}

\newcommand{\1}{{\mathbbm{1}}}

\newcommand{\nocontentsline}[3]{}
\newcommand{\tocless}[2]{\bgroup\let\addcontentsline=\nocontentsline#1{#2}\egroup}

\newif\iffigs
\figstrue

\begin{document}

\title{Experimental Quantum Computations on a Topologically Encoded Qubit}

\author{Daniel Nigg$^1$}
\altaffiliation{These authors contributed equally to this work.}
\author{Markus M\"uller$^{2}$}
\altaffiliation{These authors contributed equally to this work.}
\author{Esteban A. Martinez$^{1}$}
\author{Philipp Schindler$^{1}$}
\altaffiliation{Present address: Department of Physics, University of California, CA, USA}
\author{Markus Hennrich$^{1}$}
\author{Thomas Monz$^{1}$}
\author{Miguel Angel Martin-Delgado$^{2}$}
\author{Rainer Blatt$^{1,3}$}

\affiliation{$^1$Institut f\"ur Experimentalphysik, Universit\"at Innsbruck, Technikerstrasse 25, A--6020 Innsbruck, Austria\\
$^{2}$Departamento de F\'isica Te\'orica I, Universidad Complutense, Avenida Complutense s/n, 28040 Madrid, Spain \\
$^3$Institut f\"ur Quantenoptik und Quanteninformation der \"Osterreichischen Akademie der Wissenschaften,
Technikerstrasse 21a, A--6020 Innsbruck, Austria
}


\begin{abstract}
The construction of a quantum computer remains a fundamental
scientific and technological challenge, in particular due to
unavoidable noise. Quantum states and operations can be protected from
errors using protocols for fault-tolerant quantum computing (FTQC).
Here we present a step towards this by implementing a quantum error
correcting code, encoding one qubit in entangled states distributed
over 7 trapped-ion qubits. We demonstrate the capability of the code
to detect one bit flip, phase flip or a combined error of both,
regardless on which of the qubits they occur. Furthermore, we apply
combinations of the entire set of logical single-qubit Clifford gates
on the encoded qubit to explore its computational capabilities. The
implemented 7-qubit code is the first realization of a complete
Calderbank-Shor-Steane (CSS) code and constitutes a central building
block for FTQC schemes based on concatenated elementary quantum
codes. It also represents the smallest fully functional instance of
the color code, opening a route towards topological FTQC.
\end{abstract}

\maketitle



A fully-fledged quantum computer can be used to efficiently solve notoriously difficult problems, such as factoring large numbers or simulating the dynamics of many-body  quantum systems \cite{nielsen-book}. Enormous technological progress has enabled the implementation of small-scale prototype quantum computing devices on diverse physical platforms \cite{ladd-nature-464-45}. Similarly, sophisticated FTQC techniques have been developed, which aim at the systematic correction of errors that dynamically occur during storage and manipulation of quantum information \cite{shor-pra-52-2493(R), calderbank-pra-54-1098, steane-prl-77-793}. For quantum error correction,
CSS codes \cite{calderbank-pra-54-1098,steane-prl-77-793} offer the advantage that they allow one to independently detect and correct bit and phase flip errors, as well as combinations thereof. Furthermore, quantum information processing is substantially facilitated in quantum codes, in which logical operations on encoded qubits are realized by the bitwise application of the corresponding operations to the underlying physical qubits, i.e.~in a transversal way. This property prevents uncontrolled propagation of errors through the quantum hardware, which in turn is essential to enter the FTQC regime \cite{nielsen-book}. Ultimately, reliable quantum memories and arbitrarily long quantum computations are predicted to become feasible for appropriately designed quantum codes, once all elementary operations are realized in a fault-tolerant way and with sufficiently low error rates \cite{shor96a, preskill-review}.

To date topological quantum computing (TQC) stands up as the most promising and realistic approach towards FTQC: here, encoding of quantum information in global properties of a many-particle system provides protection against noise sources that act locally on individual or small sets of qubits \cite{kitaev-annalsphys-303-2}.
Most prominently, TQC offers highly competitive error thresholds as high as 1\% per operation
\cite{dennis-j-mat-phys-43-4452, raussendorf-prl-98-190504, katzgraber-prl-103-090501, wang-quant-inf-comp-10-780}, which is within reach of current experimental capabilities \cite{benhelm-nphys-4-463, harty-arXiv:1403.1524, barends-arXiv:1402.4848} and typically about two orders of magnitude larger than in schemes using concatenated quantum codes \cite{preskill-review}.

Within TQC, topological color codes \cite{bombin-prl-97-180501,bombin-prl-98-160502} offer the distinctive feature that the entire group of Clifford gate operations can be implemented transversally \cite{nielsen-book}. This versatile set of operations directly enables protocols for quantum distillation of entanglement, quantum teleportation and dense coding with topological protection \cite{bombin-prl-97-180501}. Moreover, a universal gate set, enabling the implementation of arbitrary quantum algorithms, can be achieved by complementing the Clifford operations with a single non-Clifford gate \cite{nielsen-book}. For color codes in two-dimensional (2D) architectures \cite{bombin-prl-97-180501}, such an additional gate can be realized by a technique known as magic-state injection \cite{bravyi-pra-71-022316}. Remarkably, this method is not needed in 3D color codes that allow one to implement a universal gate set using exclusively transversal operations \cite{bombin-prl-98-160502}.

\begin{figure*}[ht]
\begin{center}
\begin{minipage}[l]{1\textwidth}
	\begingroup
	\parfillskip=0pt
	\begin{minipage}[p]{11.7cm}
        	\begin{center}
	\includegraphics[width=11.7cm]{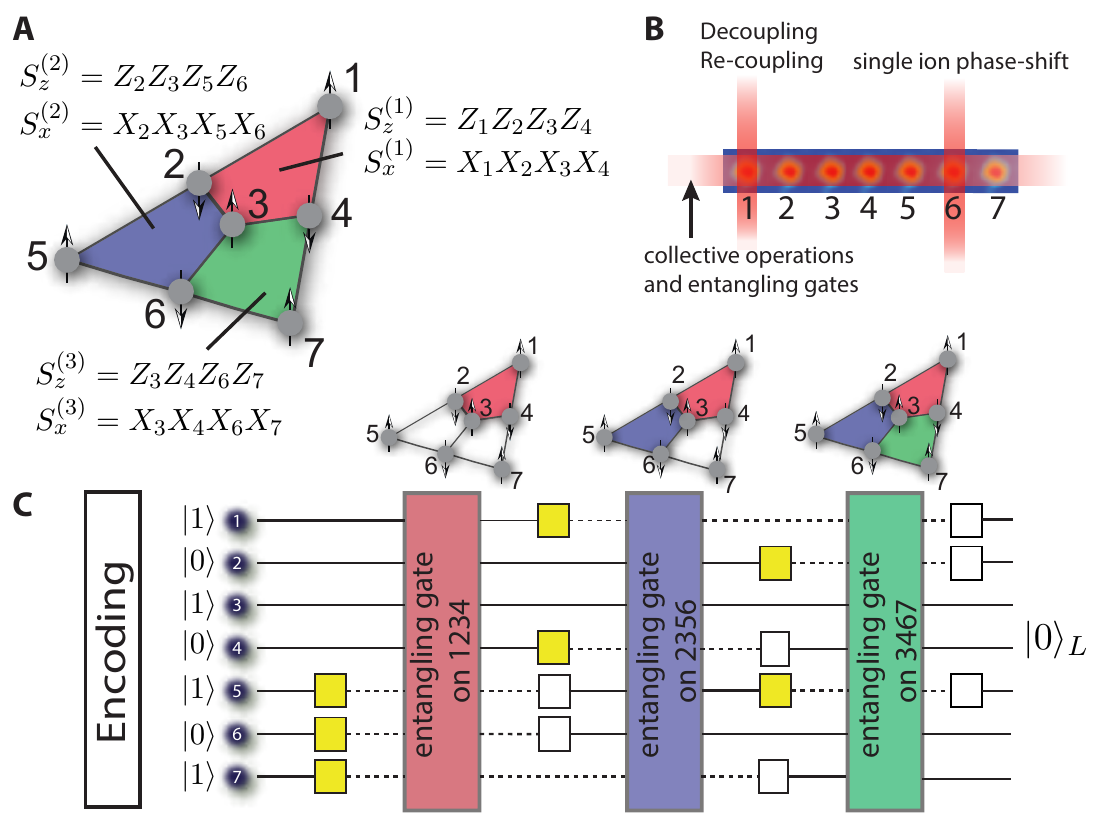}
	\end{center}
        \vspace{-4mm}
        \end{minipage}%
	\begin{minipage}[p]{6.1cm}
	\vspace{2.5mm}
	\justifying
         \textbf{Fig.~1.~The topologically-encoded qubit, its physical implementation and initialization.}
    \textbf{(A)} One logical qubit is embedded in seven physical qubits forming a 2D
    tringular planar code structure of three plaquettes. The code space is defined
    via six stabilizer operators $S_x^{(i)}$ and $S_z^{(i)}$, each acting on a plaquette which involves four physical qubits.
    \textbf{(B)} For the physical realization, qubits
    are encoded in electronic states of a linear string of ions.
    These ion qubits can be manipulated via laser-interactions that realize
    a universal gate set consisting of single-ion phase shifts,
    collective operations and a collective entangling gate, which is complemented by a single-ion spectroscopic de- and
    recoupling technique \cite{schindler-njp-15-123012}. \textbf{(C)} Encoding
    of the logical qubit is achieved by coherently mapping the input state $\ket{1010101}$ onto the logical state $\ket{0}_L$, using a quantum circuit that combines plaquette-wise entangling operations with de- and recoupling pulses (yellow and white squares, respectively) \cite{si}. Dashed (solid) lines denote decoupled or inactive (recoupled or active) qubits.
	\end{minipage}%
	\par
	\endgroup
\end{minipage}
	\vspace{-2mm}
\end{center}
\end{figure*}

Previous experiments have demonstrated the correction of a single type of error by the 3-qubit repetition code \cite{chiaverini-nature-432-602, schindler-science-332-1059, reed-nature-482-382}, correction of bit and phase flip errors by the non-CSS-type 5-qubit code in NMR systems \cite{knill-prl-86-5811,zhang-prl-109-100503}, as well as elements of topological error correction in the framework of measurement-based quantum computation \cite{yao-nature-482-489}. In this work we report on the first demonstration of a quantum error correcting 7-qubit CSS code \cite{steane-prl-77-793} which is equivalent to the smallest instance of a 2D topological color code \cite{bombin-prl-97-180501}. In our experiments performed with 7 trapped-ion qubits we encode and characterize a single logical qubit. Moreover, we implement in a transversal way the entire set of logical single-qubit Clifford gate operations on the encoded qubit. Finally, the application of multiple logical gate operations for the first time realizes a quantum computation on a fully-correctable encoded qubit.

Two-dimensional color codes are topological quantum error-correcting codes that are constructed on underlying 2D lattices \cite{bombin-prl-97-180501} for which (i) three links meet at each vertex and (ii) three different colors are sufficient to assign color to all polygons (plaquettes) of the lattice such that no adjacent plaquettes sharing a link are of the same color. The smallest, fully functional 2D color code involves seven qubits and is shown in Fig.~1A. The system consists of a triangular, planar code structure formed by three adjoined plaquettes with one physical qubit placed at each vertex. Color codes are stabilizer quantum codes \cite{nielsen-book}, which are defined by a set of commuting, so-called stabilizer operators $\{S_i\}$, each having eigenvalues +1 or -1. More precisely, the code space hosting logical or encoded quantum states $\ket{\psi}_L$
is fixed as the simultaneous eigenspace of eigenvalue +1 of all stabilizers, $S_i \ket{\psi}_L = + \ket{\psi}_L$ \cite{gottesman-pra-54-1862, nielsen-book}. In color codes, there are two stabilizer operators associated with each plaquette, which for the seven-qubit color code (Fig.~1A) results in the set of four-qubit $X$ and $Z$-type operators
\begin{align}
S_x^{(1)} &=  X_1 X_2 X_3 X_4, \qquad S_z^{(1)} = Z_1 Z_2 Z_3 Z_4, \nonumber \\
\label{eq:stabilizer_set}
S_x^{(2)} &=  X_2 X_3 X_5 X_6, \qquad S_z^{(2)} = Z_2 Z_3 Z_5 Z_6, \\
S_x^{(3)} &=  X_3 X_4 X_6 X_7, \qquad S_z^{(3)} = Z_3 Z_4 Z_6 Z_7. \nonumber
\end{align}
Here, $X_i$, $Y_i$ and $Z_i$ denote the standard Pauli matrices acting on the $i$-th physical qubit with the computational basis states $\ket{0}$ and $\ket{1}$ \cite{nielsen-book}. The stabilizers in Eq.~(\ref{eq:stabilizer_set}) impose six independent constraints on the seven physical qubits and thus define a two-dimensional code space, which allows one to encode one logical qubit. The logical basis states $\ket{0}_L$ and $\ket{1}_L$ spanning the code space are entangled 7-qubit states and given as the eigenstates of the logical operator $Z_L = Z_1 Z_2 Z_3 Z_4 Z_5 Z_6 Z_7$, where $Z_L \ket{0}_L = \ket{0}_L$ and $Z_L \ket{1}_L = - \ket{1}_L$ \cite{si, bombin-prl-97-180501}.


For the physical realization of a topologically encoded qubit, we
store seven $^{40}$Ca$^+$ ions in a linear Paul trap. Each ion hosts a
physical qubit, which is encoded in (meta)stable electronic states \cite{schindler-njp-15-123012}.
Within our setup, schematically shown in Fig.~1B, we realize a high-fidelity universal set of quantum
operations consisting of single-ion phase shifts, collective
rotations, and a collective entangling gate. This set of operations is complemented by a single-ion spectroscopic decoupling technique that enables the collective
entangling operation to act only on subsets of qubits \cite{si}.


\begin{figure*}[ht]
\begin{center}
\begin{minipage}[l]{1\textwidth}
	\begingroup
	\parfillskip=0pt
	\begin{minipage}[p]{11cm}
        	\begin{center}
	\includegraphics[width=11cm]{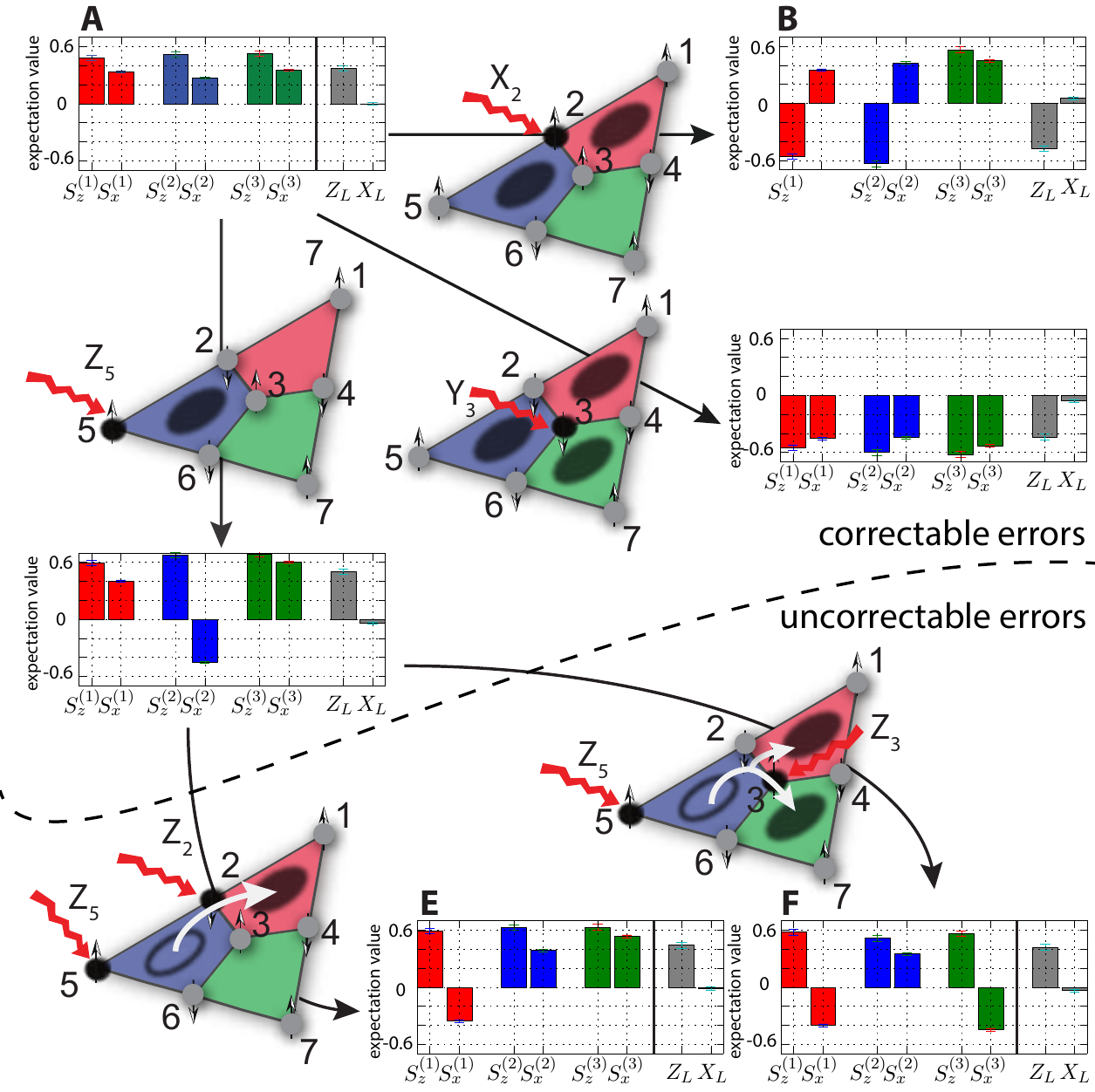}
	\end{center}
        \vspace{-4mm}
        \end{minipage}%
	\hfill
	\begin{minipage}[p]{6.5cm}
	\justifying
         \textbf{Fig.~2.~Effect of arbitrary single-qubit errors on the encoded logical qubit.}
(\textbf{A}) The initial logical state $\ket{0}_L$, prior to the occurrence of single-qubit errors,
is reflected (i) by the error syndrome, in which all six $S_x^{(i)}$ and $S_z^{(i)}$ stabilizers are positive-valued, and (ii) by a positive (vanishing) expectation value of the logical operator $Z_L$ ($X_L$). (\textbf{B}) A bit flip error (red wiggled arrow) on qubit 2 (marked in black) affects the blue and red plaquettes (visualized by grey-shaded circles) and manifests itself by negative $S_z^{(1)}$ and $S_z^{(2)}$ expectation values and a $Z_L$ sign flip. (\textbf{C}) A $Z_5$ phase flip error only affects the blue plaquette and results in a sign flip of $S_x^{(2)}$. (\textbf{D}) A $Y_3$ error -- equivalent to a combined $X_3$ and $Z_3$ error --
affects all three plaquettes and induces a sign change in all six stabilizers and $Z_L$. Double-error events, such as a $Z_5$ phase flip (Fig.~2C), followed by a $Z_2$ (\textbf{E}) or a $Z_3$ error (\textbf{F}) result in an incorrect assignment of physical errors, as the detected stabilizer patterns are indistinguishable from single-error syndromes -- here, the ones induced by a $Z_1$ (Fig.~2E) or a $Z_4$ (Fig.~2F) error \cite{si}. In the correction process, this eventually results in a logical error -- here a $Z_L$ phase flip error. Stabilizer violations can under subsequent errors hop (white non-wiggled arrow) to an adjacent plaquette, as in Fig.~2E, where the violation disappears (open grey circle) from the blue and reappears on the red plaquette. Alternatively (Fig.~2F), they can disappear (from the blue plaquette), split up (white branched arrow) and reappear on two neighboring plaquettes (red and green). This rich dynamical behavior of stabilizer violations is a characteristic signature of the topological order in color codes \cite{bombin-prl-97-180501, si}.
	\end{minipage}%
	\par
	\endgroup
\end{minipage}
\end{center}
\end{figure*}



We realize the initial preparation of the logical state $\ket{0}_L$ (encoding) deterministically
by applying the quantum circuit shown
in Fig.~1C to the seven-ion system. As a first step, we prepare the system in the
product state $\ket{1010101}$, which satisfies the required three $\langle
S_z^{(i)} \rangle=+1$ and $\langle Z_L \rangle =+1$ conditions. In three subsequent
steps, we apply plaquette-wise entangling operations
to also satisfy the three $\langle S_x^{(i)} \rangle=+1$ constraints. For each
step, we spectroscopically decouple three of the seven physical qubits prior to the application
of the collective entangling gate, to subsequently create GHZ-like
entanglement only between the four qubits belonging to one plaquette of the triangular code.
Here, the creation of entanglement of four out of seven qubits is achieved with a fidelity of 88.8(5)\%.

The entire sequence for encoding involves three collective entangling
gates and 108 local single-qubit rotations~\cite{si}.
The quantum state fidelity of the system in state $\ket{0}_L$ is exactly determined
from measurements of 128 Pauli operators, and yields 32.7(8)\%.
This value surpasses the threshold value of 25\% (by more than 9 standard deviations),
above which genuine six-qubit entanglement is witnessed, thereby clearly
indicating the mutual entanglement of all three plaquettes of the code~\cite{si}.

The quality of the created logical state $\ket{0}_L$, as shown in Fig.~2A, is governed by
two factors: (a) the overlap of the created state with the code space;
and (b) the accordance of the experimental state \textit{within the code space} with the target encoded state,
which is related to the expectation values of the logical operator $Z_L$.
Residual populations outside the code-space are
indicated by deviations of the six stabilizer expectation values, in
our case on average 0.48(2), from the ideal value of +1. A more
detailed analysis shows that within the
code-space, the fidelity between experimental and target state
is as high as 95(2)\%, whereas the expectation value
of $\langle Z_L \rangle = 0.38(3)$ and
the overall fidelity between the experimentally realized
and the ideal state $\ket{0}_L$ are currently limited by
the overlap with the code-space of 34(1)\%~\cite{si}.

It is a hallmark feature of topologically ordered states that these cannot be characterized by local order parameters,
but only reveal their topological quantum order in global system properties \cite{kitaev-annalsphys-303-2, si}. We experimentally confirm this intriguing
characteristics for the topologically encoded 7-qubit system in state $\ket{1}_L$ by measuring all subsets of reduced two-qubit density matrices, which yield an average Uhlmann-fidelity of 98.3(2)\% with the two-qubit completely-mixed state, clearly showing the absence of any single- and two-qubit correlations. On the contrary, we observe the presence of global quantum order, as signaled for the system size at hand by non-vanishing three-qubit correlations $\langle Z_1 Z_4 Z_7 \rangle = -0.46(6)$ \cite{si}.


\begin{figure*}[ht]
\begin{center}
\begin{minipage}[l]{1\textwidth}
	\begingroup
	\parfillskip=0pt
       	\begin{center}
	\includegraphics[width=1\textwidth]{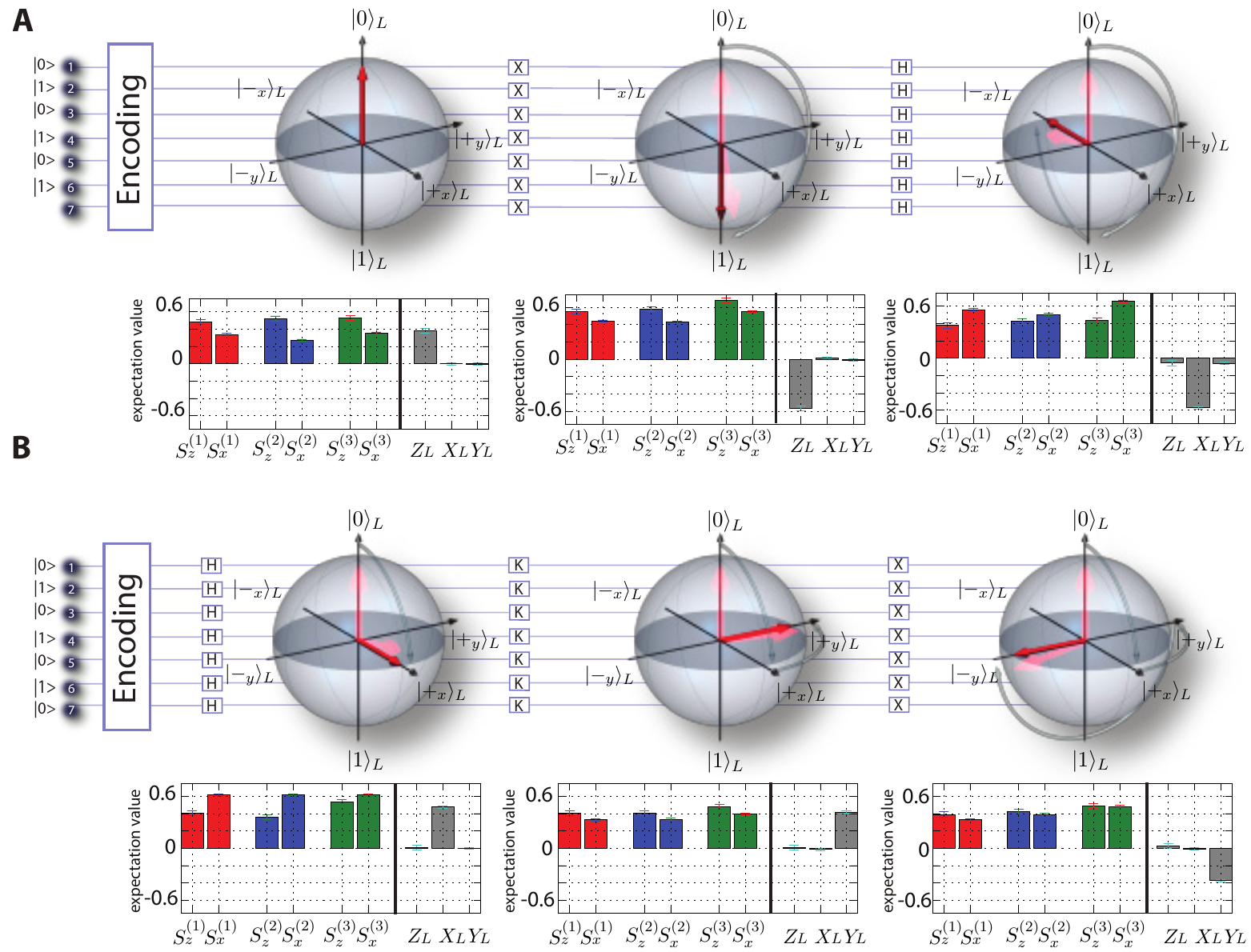}
	\end{center}
	\justifying
         \textbf{Fig.~3.~}\textbf{Single qubit Clifford gate operations applied on a logical
  encoded qubit.} Starting from the logical $\ket{0}_L$
  state, sequences of logical Clifford gate operations $\{X_L,H_L\}$
  in \textbf{(A)} and $\{H_L,K_L,X_L\}$ in \textbf{(B)}
  are applied consecutively in a transversal way (i.e. bit-wise) to realize all six cardinal states \{$\ket{0}_L,\ket{1}_L, \ket{-_x}_L, \ket{+_x}_L, \ket{+_{y}}_L, \ket{-_{y}}_L$\} of the logical space of the topologically encoded qubit. The dynamics under the applied gate operations
  is illustrated by rotations of the Bloch-vector (red arrow) on the logical Bloch-sphere as well as by the circuit diagram in the background. Each of the created logical states is characterized by the measured pattern of $S_x^{(i)}$ and $S_z^{(i)}$ stabilizers and the logical Bloch vector, with the three components given by the expectation values of the logical operators $X_L$, $Y_Z$ and $Z_L$. The orientation of the logical Bloch vector changes as expected under the logical gate operations.
	\par
	\endgroup
\end{minipage}
	\vspace{-2mm}
\end{center}
\end{figure*}


\begin{figure*}[ht]
\begin{center}
\begin{minipage}[l]{1\textwidth}
	\begingroup
	\parfillskip=0pt
       	\begin{center}
	\includegraphics[width=1\textwidth]{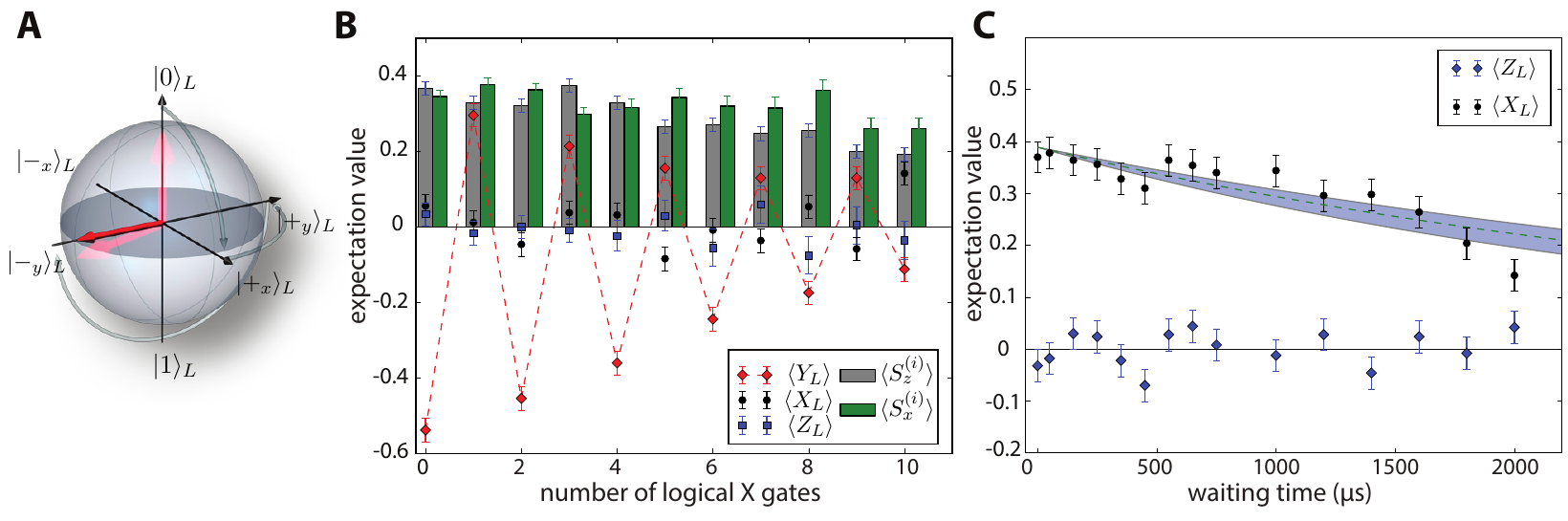}
	\end{center}
	\justifying
  \textbf{Fig.~4.~Repetitive application of logical quantum gate
    operations:} \textbf{(A)} Preparation of the $\ket{-_y}_L$ state
  by applying a $H_L$, $K_L$ and $X_L$ gate operation on the qubit
  initially prepared in the $\ket{0}_L$ state. \textbf{(B)}
  Subsequently, flips between the logical $\ket{+_y}_L$ and
  $\ket{-_y}_L$ states are induced by consecutively applying logical
  $X_L$ gate operations up to 10 times. The sign flip of the $Y_L $
  expectation value (red diamonds) after each step signals clearly the
  induced flips of the logical Bloch vector, whereas the expectation
  values of $Z_L$ (blue squares) and $X_L$ (black circles) are close
  to zero as expected (the average of \{$\langle Z_L \rangle$,
  $\langle X_L \rangle$\} yields \{$0.01(1),-0.01(1)$\} \cite{si}).
  Average $S_z^{(i)}$ ($S_x^{(i)}$) stabilizer expectation values
  after each $X_L$ gate are shown as grey (green) bars. \textbf{(C)}
  Characterization of the coherence of the logical qubit, initially
  prepared in the $X_L$ eigenstate $\ket{+_x}_L$. A measurement of the
  decay of the $X_L$ expectation value (black circles) as a function
  of time yields a $1/e$-time of 3.6(6)ms \cite{si}, while the $Z_L$
  expectation value (blue diamonds) remains zero as expected (on
  average 0.001(8) \cite{si}).
	\par
	\endgroup
\end{minipage}
	\vspace{-2mm}
\end{center}
\end{figure*}


We next study the error correction properties of the encoded qubit. Single-qubit errors lead the system out of the
logical code space and manifest themselves as stabilizer eigenstates of eigenvalue -1
(stabilizer violations) associated to one or several plaquettes.
We use the available gate set to coherently induce all single-qubit errors
on the encoded state $\ket{0}_L$
and record the induced error syndromes provided by the
characteristic pattern of six stabilizer expectation values \cite{si} -- see
Fig.~2 for a selection. The experimental data
clearly reveals the CSS character of the quantum code:
Starting in $\ket{0}_L$ (Fig.~2A), we observe that single-qubit $X$ ($Z)$ errors
manifest themselves as violations of $Z$-type ($X$-type) stabilizers
only (Figs.~2B and C); the
effect of single-qubit $Y$ errors is equivalent to a combined
$X$ and $Z$ error and is indeed signaled by the simultaneous violation of the
corresponding $X$ and $Z$-type stabilizers
(Fig.~2D). For our experimental statistical uncertainties,
the measured characteristic error syndromes can be perfectly assigned to the
underlying induced single-qubit error \cite{si}.
Figures~2E and F show data where
the code has been exposed to two single-qubit errors, whose correction
exceeds the capabilities of the 7-qubit code \cite{steane-prl-77-793, bombin-prl-97-180501}.


In topological color codes, quantum information is processed by logical gate operations acting directly within the code space \cite{bombin-prl-97-180501, bombin-prl-98-160502}. The entire group of logical Clifford gates, generated by the elementary gate operations $Z$, $X$, the Hadamard $H$ and the phase gate $K$,
\begin{equation}
H = \frac{1}{\sqrt{2}} \left(\begin{array}{cc}
1 & 1 \\
1 & -1
\end{array}\right), \qquad K = \left(\begin{array}{cc}
1 & 0 \\
0 & i
\end{array}\right),
\end{equation}
as well as the C-NOT gate for registers containing several logical qubits, can be realized transversally. We implement the logical Clifford gates $Z_L = Z_1Z_2Z_3Z_4Z_5Z_6Z_7$, $X_L = X_1X_2X_3X_4X_5X_6X_7$, $H_L = H_1H_2H_3H_4H_5H_6H_7$ and $K_L = K_1K_2K_3K_4K_5K_6K_7$ by local operations which we realize by a combination of single-ion and collective rotations~\cite{si}, as offered by our setup shown in Fig.~1B. After initialization of the logical qubit in the state $\ket{0}_L$, we prepare all six eigenstates of the logical operators $X_L$, $Y_L$ and $Z_L$, which requires quantum circuits (see Figs.~3A and B) consisting of up to three elementary logical Clifford gate operations. We clearly observe that the encoded qubit evolves as expected under the logical gate operations, as indicated by the characteristic changes in the pattern of $X_L$, $Y_L$ and $Z_L$ expectation values, and corroborated by quantum state fidelities within the code space of \{95(2), 85(3), 87(2)\}\% of the experimental logical states $\{ \ket{0}_L$, $\ket{1}_L$ and $\ket{+_x}_L = (\ket{0}_L + \ket{1}_L)/\sqrt{2}$\} with the expected ideal states~\cite{si}. Furthermore, the data shows that the average values of the six stabilizers for each logical state is uncorrelated from the number of logical gates applied to prepare them (up to three) \cite{si}. This indicates that currently imperfections in the initial encoding process dominate over the influence of additional errors induced by subsequently applied short sequences of few Clifford gate operations. This is an indication of the high performance of the transversal (i.e.~bitwise) logical Clifford gate operations, as compared to the encoding which involves numerous multi-qubit as well as single-ion entangling operations. In fact, this behavior is confirmed by measured overlap with the code space of \{34(1), 33(2), 39(2)\}\% for the states $\{ \ket{0}_L$, $\ket{1}_L$ and $\ket{+_x}_L \}$ \cite{si}.

We further explore the computational capabilities of the encoded qubit
by executing a longer encoded quantum computation, which consists of
up to 10 logical $X_L$ gate operations applied to the system initially
prepared in $\ket{-_y}_L$ (see Fig.~4A). Our data (see Fig.~4B)
clearly reveals the expected flips of the logical qubit between the
$Y_L$ eigenstates, as witnessed by alternating (vanishing) expectation
values of the logical $Y_L$ ($X_L$, $Z_L$) operator, accompanied by a
moderate decay of the average stabilizer expectation values of only
3.8(5)\% per logical gate operation \cite{si}. For a separate study of
the decoherence properties of the encoded qubit, we prepare the
logical superposition state $\ket{+_x}_L$ and measure the decay
dynamics of the expectation values of the stabilizers and logical
$X_L$ and $Z_L$ operators, see Fig.~4C. We find that logical
coherences $\langle X_L \rangle$ decay on a time scale that is about
one order of magnitude larger than the duration of 10 $X_L$
gates~\cite{si}. Thus, we conclude that the performance of the encoded
quantum computation of Fig.~4B is currently not limited by the
life-time of the encoded logical states, but is instead determined by
imperfections in the implementation of the logical Clifford gate
operations, whose quality is in quantitative agreement with the
fidelities of the single-ion and collective local operations in our
setup \cite{si}.


In this work, using 7 trapped-ion qubits, we have implemented and experimentally characterized a quantum error correcting 7-qubit CSS code \cite{steane-prl-77-793}, which represents the smallest 2D topological color code \cite{bombin-prl-97-180501}, hosting one encoded logical qubit. Moreover, we have realized encoded quantum computations by applying sequences of logical Clifford gate operations on the fully-correctable, topologically encoded qubit. Near future objectives, aiming at an extension of the demonstrated quantum information processing capabilities, include the implementation of a non-Clifford gate, using one additional (unprotected) ancillary qubit, towards universal encoded quantum computation \cite{bravyi-pra-71-022316, nielsen-book}. Furthermore, repetitive application of complete error correction cycles is achievable by incorporating previously demonstrated measurement and feedback techniques \cite{barreiro-nature-470-486} for quantum non-demolition (QND) readout of stabilizers via ancillary qubits, towards the goal of keeping an encoded qubit alive. The demonstrated concepts can be adapted to 2D ion-trap arrays \cite{muir-arXiv:1402.0791} as well as other scalable architectures, ranging from optical, atomic and molecular to solid-state systems \cite{ladd-nature-464-45}. The continuing technological development of these platforms promises fault-tolerant operating of larger numbers of error-resistant, topologically protected qubits, and thereby marks a technologically challenging, though clear path towards the realization of FTQC.

\section*{Acknowledgments}
We gratefully acknowledge support by the Spanish MICINN grant
FIS2009-10061, FIS2012-33152, the CAM research consortium QUITEMAD
S2009-ESP-1594, the European Commission PICC: FP7 2007-2013, Grant No.
249958, the integrated project SIQS (grant No. 600645), the UCM-BS
grant GICC-910758, and by the Austrian Science Fund (FWF), through the
SFB FoQus (FWF Project No. F4002-N16), as well as the Institut f\"ur
Quanteninformation GmbH. This research was supported by the U.S. Army
Research Office through grant W911NF-14-1-0103.


\section*{Author Contributions}
M.M., D.N., E.A.M., P.S., T.M., M.A.M.D. and R.B. developed the research and conceived the experiment; D.N., E.A.M. and M.M. performed the experiments; D.N., E.A.M. and T.M. analyzed the data; D.N., E.A.M., T.M., P.S., M.H. and R.B. contributed to the experimental set-up; M.M., D.N., T.M. and M.A.M.D. wrote the manuscript; all authors contributed to the discussion of the results and manuscript.

\vspace{1cm}
\appendix


\begin{center}\textbf{APPENDICES}\end{center}

\vspace{3mm}This appendix provides theoretical background information and experimental details of the material discussed in the main text. In addition, it contains experimental data which due to space limitations has not been discussed in the main text of the paper: this includes a complete measured syndrome table containing all 21 single-qubit error syndromes (see Fig.~\ref{fig:syndrome}), and quantum state tomography data confirming the absence (presence) of local (global) quantum order in the experimental 7-qubit state (see Sec.~\ref{sec:SI_topological_order} and Fig.~\ref{fig:suppl_tomo}).

In Sec.~\ref{sec:SI_experimental_system} we provide details on our trapped-ion quantum computing setup and the experimental tools used in this work. Details on the initialization procedure (encoding) are provided in Sec.~\ref{sec:SI_encoding}. In Sec.~\ref{sec:SI_state_characterization} we discuss the methods used to quantitatively characterize the experimentally generated states, which include tools to determine the quantum state fidelities and overlap with the code space, to witness multi-partite entanglement properties, and to study aspects of topological quantum order in the experimentally generated state. Section~\ref{sec:SI_syndrome} provides details on the experimental study of the error detection capabilities of the 7-qubit code. Finally, in Sec.~\ref{sec:SI_Clifford_gates} details on the implementation and characterization of sequences of logical Clifford gate operations on the encoded qubit are discussed.


\begin{table*}
  \centering
  \begin{tabular}{|c|c|c|c|c|c|c|}
    \hline
     & Nr.~of & Nr.~of global & Nr.~of ac- & Nr.~of & Total  & Described  \\
    Algorithm & MS gates & rotations $R$ & Stark  & addressed  & number of & in section\\
       & & & shifts $S_Z$ & resonant pulses  & operations & \\
    \hline
    Decoupling / Recoupling of population in $\ket{0}$ & 0 & 0 & 2 & 4 & 6 & IB \\
    Decoupling / Recoupling of population in $\ket{1}$ & 0 & 0 & 1 & 2 & 3 & IB\\
    Complete decoupling / Recoupling of a physical qubit & 0 & 0 & 3 & 6 & 9 & IB\\
        \hline
    Encoding: preparation of the logical $\ket{1}_{L}$ state & 3 & 1 & 38 & 70 & 112 & II \\
       \hline
    Logical Z gate & 0 & 0 & 7 & 0 & 7 & V \\
    Logical X gate & 0 & 0 & 0 & 1 & 1 & V \\
    Logical Hadamard gate H & 1 & 7 & 0 & 0 & 8 & V \\
    Logical phase gate K & 0 & 0 & 7 & 0 & 7 & V\\
    \hline
    Preparation of the logical $\ket{0}_{L}$ state & 3 & 0 & 38 & 70 & 111 & V\\
    Preparation of the logical $\ket{+_{x}}=\ket{0}_{L} +  \ket{1}_{L}$ state & 3 & 1 & 45 & 70 & 118 & V\\
    Preparation of the logical $\ket{-_{x}}=\ket{0}_{L} - \ket{1}_{L}$ state & 3 & 2 & 45 & 70 & 119 & V\\
    Preparation of the logical $\ket{+_{y}}=\ket{0}_{L} + i \ket{1}_{L}$ state & 3 & 1 & 52 & 70 & 125 & V\\
    Preparation of the logical $\ket{-_{y}}=\ket{0}_{L} - i \ket{1}_{L}$ state & 3 & 2 & 52 & 70 & 126 & V\\
    \hline
    Encoding and up to 13 Clifford gates & 3 & 14 & 52 & 70 & 136 & VI\\
    \hline
  \end{tabular}
  \caption{Overview of the building blocks $\{$MS gate, global rotations (X,Y), ac-Stark shifts $S_{Z}$, addressed resonant pulses$\}$ used for the individual parts of the encoding sequence, the preparation of the 6 logical states and the logical Clifford gate operations. For the preparation of the logical
$|0\rangle$ and $|1\rangle$ state, 3 entangling MS gates, one global rotation for the logical $|1\rangle$ state, 38 ac-stark operations and 70 addressed resonant pulses
for the de- and recoupling of the individual ions are required, which in total correspond to 111 (112) gate operations. For the preparation of the logical states $\{\ket{+_{x}},\ket{-_{x}}, \ket{+_{y}},\ket{-_{y}}\}$ in total $\{118,119,125,126\}$ gate operations are used, whereas for the demonstration of multiple Clifford gate operations we apply up to 136 gate operations in total.
}
  \label{tab:Resources}
\end{table*}

\section{Experimental system and techniques}
\label{sec:SI_experimental_system}

This section provides a more detailed description of the experimental setup, especially the
coherent gate operations as well as the spectroscopic decoupling technique.

All experiments described in this work are performed using a linear string of 7 $^{40}$Ca$^+$ ions
confined in a Paul trap \cite{s-schindler-njp-15-123012}. The (physical) qubit is encoded and manipulated on the transition between
the two electronic states $S_{1/2}(m_{j}=-1/2) = |1\rangle$
$\rightarrow$ $D_{5/2}(m_{j}=-5/2) = |0\rangle$, the latter having a radiative lifetime of about $1s$. One generic experimental
cycle consists of the following sequence: (i) Initialization of all qubits in the electronic $S_{1/2}(m_{j}=-1/2) = |1\rangle$
ground state by optical pumping. (ii) Cooling the ion string to the motional ground state of the center of mass mode.
(iii) Coherent operations on the $S_{1/2}(m_{j}=-1/2) \rightarrow D_{5/2}(m_{j}=-1/2)$ transition
by a narrow linewidth laser at 729\,nm. (iv) Detection of the final state by fluorescence measurements, involving an electron shelving
technique using a laser beam at 397\,nm illuminating the whole ion string. For more details on the
individual techniques, see the following subsections and Ref.~\cite{s-schindler-njp-15-123012}.

\subsection{Coherent gate operations}
\label{sec:SI_coherent_gates}
In our experimental setup, two different laser beams at 729\,nm are used to perform
coherent operations on the qubit transition. A spatially wide beam illuminating the whole
ion string realizes collective operations of the form
\begin{equation}
U(\theta,\phi) =
\exp \left(-i\frac{\theta}{2}\sum_{i}\left[\sin(\phi)Y_i+\cos(\phi)X_i\right] \right)
\end{equation}
and a M$\o$lmer-S$\o$rensen-type entangling operation \cite{s-molmer-prl-82-1835, s-roos-njp-10-013002}
\begin{equation}
MS(\theta,\phi) =
\exp \left(-i\frac{\theta}{4}\left[\sum_{i}\sin(\phi)Y_i+\cos(\phi)X_i\right]^{2} \right).
\end{equation}
Since the angle of incidence with of the global beam with respect to the ion string is approximately 22.5$^{\circ}$,
the beam is shaped elliptically to illuminate all ions equally. The beam ellipticity
is about 1:5 with a beam waist of 100\,$\mu m$ in horizontal direction, which leads to
an inhomogeneity of the coupling strength along the 7 ion string, measured by the Rabi-frequency, of
about 1$\%$. In addition, a beam perpendicular to the ion string, which is focused to a beam waist
of about 1.5$\mu m$ is used to perform single qubit rotations $U_{Z}^{(i)}(\theta) =
\exp(-i\frac{\theta}{2}Z_i)$ on the $i$-th ion. This operation, corresponding
to a rotation around the $Z$-axis in the Bloch-sphere picture, is carried out
by detuning the laser beam about 20\,MHz from the qubit transition, which effectively
induces an AC-Stark shift \cite{s-schindler-njp-15-123012}. The combination of the described
gate operations realizes a universal set of gate operations \cite{s-nebendahl-pra-79-012312, s-mueller-njp-13-085007}.
Alternatively, the addressing beam is also capable of inducing resonant operations of the form $U^{(i)}(\theta,\phi) =
\exp(-i\frac{\theta}{2}\{\sin(\phi)Y_i+\cos(\phi)X_i\})$.

\subsection{Spectroscopic decoupling and recoupling of ions}
\label{sec:SI_decoupling}

For the experimental initialization of the encoded qubit in the logical state $|0\rangle_{L}$, the main resource
is an entangling operation acting on a subset of four out of seven qubits. In Figure~\ref{fig:hiding}A the circuit diagram of the sequence used for the encoding
is shown. In the first step an entangling operation $MS(\pi/2,0)$
is to be applied to ions 1, 2, 3, 4 without affecting the ions 5, 6 and 7. However, as the laser beam of the entangling MS gate operation illuminates the entire ion string, an entangling operation on a subset of ions can be achieved in two different ways: One can resort to refocusing techniques, as originally pioneered in NMR systems \cite{s-vandersypen-rmp-76-1037}. Here, partially entangling \textit{global} entangling MS gate operations are interspersed with single-qubit AC Stark shifts, which eventually lead to an effective decoupling of (subsets of) ions from the entangling dynamics of the remaining ions (see e.g.~\cite{s-mueller-njp-13-085007} for more information). However, in the present case, such decoupling of 3 ions, using refocusing pulses, from the entangling dynamics of the remaining 4 qubits belonging to one plaquette of the quantum code, would require a large overhead in terms of (partially entangling) MS gate operations and addressed, single-ion refocusing pulses. Thus, in the present experiment we pursue an alternative approach, where we decouple ions \textit{spectroscopically} from the dynamics induced by the global MS gate operation. To this end, we coherently transfer and store the quantum state of qubits (say 5, 6 and 7) encoded in the states $S_{1/2}(m_{j}=-1/2)$ and $D_{5/2}(m_{j}=-1/2)$, which do not participate in the entangling dynamics of qubits 1, 2, 3, and 4, in a subset of the remaining (metastable) Zeeman levels.

The latter has the advantage that the entangling operation, acting only on the subset of 4 qubits, can be realized with a fidelity of 88.5(5)\%, as determined by a 4-qubit state tomography of the created 4-qubit GHZ state, $(\ket{0101} + \ket{1010})/\sqrt{2}$. Besides imperfections in the entangling operation, this value thus also includes the effect of small imperfections in the preparation of the initial product state $\ket{1010}$. The fidelity of the entangling operation on 4 out of 7 ions is substantially higher than the fidelity with which a 7-qubit GHZ state can be created by a global entangling operation (acting on all 7 ions). From population and parity measurements we estimate the fidelity for this 7-ion entangling operation to be about 84\%. In Figure~\ref{fig:hiding} B, the decoupling (DEC) steps, highlighted as colored boxes (DEC) are illustrated
by a reduced level scheme of the relevant Zeeman states. The decoupling sequence can be adapted, so that -- depending on the
internal state of the physical qubit -- only a minimal number of decoupling pulses have to be applied. For example, the decoupling of qubits 5 and 6 (blue and red box) before the first entangling operation is realized as follows: The population of qubit 5 (initially entirely in state $S_{1/2}(m_{j}=-1/2)$) is transformed to the $D_{5/2}(m_{j}=-5/2)$ state via a composite pulse sequence of 3 coherent single-qubit operations: a resonant $\pi/2$-pulse $U^{(5)}(\pi/2,\theta)$ with arbitrary, but fixed phase $\theta$ on qubit 5, followed by a $Z$ rotation $U_{Z}^{(5)}(\pi)$ and finally another resonant $U^{(5)}(\pi/2,\theta-\pi)$ rotation
with phase $\theta-\pi$. Effectively, this sequence realizes a $\pi$-flop (i.e.~complete population transfer) between the computational basis state $S_{1/2}(m_{j}=-1/2)$ and the storage state $D_{5/2}(m_{j}=-5/2)$. This population transfer could in principle
also be performed by a single resonant $U^{(5)}(\pi,\theta)$ pulse. The reason for splitting
this up into 3 addressed pulses is to minimize errors on neighboring ions while realizing the decoupling operation on the target ion: The application of the addressed laser beam inevitably leads to small residual light intensities on the neighboring ions, which should ideally be unaffected. This error $\epsilon$ can be characterized
by measuring the Rabi-frequencies of the neighboring ions and the target ion: $\epsilon = \Omega_{\text{neighbor}}/\Omega_{\text{target}}$.
For a seven-ion string with a minimal inter-ion distance of $\approx$ 3.5 $\mu m$ at a trap frequency of
about 1\,MHz, the addressing error $\epsilon$ is about 5\% on the ions located at the center of the chain. Contrary to the
resonant pulse, the addressing error of the off-resonant AC-Stark pulses scale with
$\epsilon^{2}$, since the Rabi-frequency $\Omega^{2}/4\Delta$ scales quadratically with the Rabi frequency $\Omega$, and for a given detuning $\Delta$. Therefore the effectively induced error on the neighboring ions in the 3-pulse sequence also scales in leading order as $\epsilon^{2}$, since
errors induced by the two resonant $\pi/2$-pulses cancel out due to the phase shift of $\pi$.\\

Similarly, spectroscopic decoupling of ions with population in the $D_{5/2}(m_{j}=-5/2)$ state is achieved by transfering the electronic population first to the
$S_{1/2}(m_{j}=1/2)$ ground state and subsequently to the $D_{5/2}(m_{j}=-3/2)$ state, as shown in the red box of Figure~\ref{fig:hiding} B.
This sequence requires in total 3 single-qubit operations: $U^{(6)}(\pi/2,\theta)$, $U_{Z}^{(6)}(\pi)$ and $U^{(6)}(\pi/2,\theta-\pi)$
on the $D_{5/2}(m_{j}=-5/2) \rightarrow S_{1/2}$ transition, and a similar 3-pulse sequence on the $S_{1/2}(m_{j}=1/2) \rightarrow D_{5/2}(m_{j}=-3/2)$ transition, respectively.
Therefore, decoupling and also recoupling (REC) of one qubit with populations in (and coherences between) both computational basis states requires in total 9 single-qubit operations with a pulse length of about 10\,$\mu s$ per pulse (see green box in of Figure~\ref{fig:hiding} B).

\section{Details on encoding of the logical qubit}
\label{sec:SI_encoding}

\begin{figure*}[t!]
\center
\includegraphics[scale=0.85]{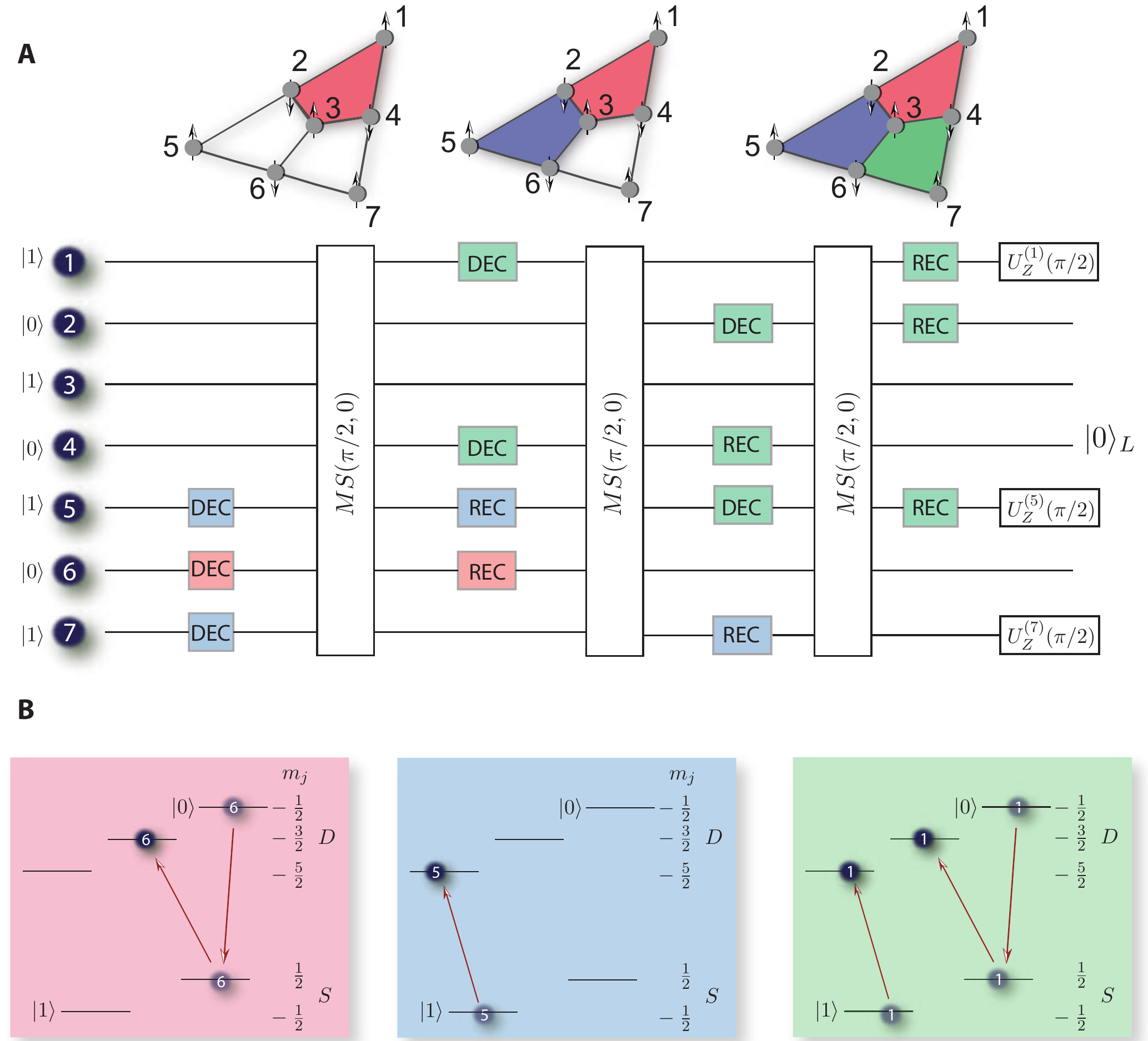}
\caption{\textbf{Initialization of the logical qubit:} (\textbf{A}) Circuit diagram of the encoding step - preparation of the 7-qubit system in the
logical state $|0\rangle_{L}$. The encoding consists of three steps, where consecutively all three colored (red, green, blue) plaquettes are
prepared by applying an entangling MS gate on the subset of qubits involved in the plaquette under consideration (e.g. qubits 1, 2, 3 and 4 for the plaquette), starting in the initial product state $|1010101\rangle$. (\textbf{B}) We achieve that qubits, which do not participate in a particular plaquette, are not affected by the entangling
operation by means of spectroscopic decoupling pulses (DEC). Here, their quantum state is stored in additional electronic levels, by coherently transferring the electronic population of the $S_{1/2}(m_{j}=-1/2)$ and $D_{5/2}(m_{j}=-1/2)$ qubit states to the Zeeman``storage" states $\{D_{5/2}(m_{j}=-5/2),D_{5/2}(m_{j}=-3/2)\}$ by a sequence
of up to 9 addressed single qubit rotations. The working principle of the decoupling and recoupling (REC) pulses is illustrated in (B) and described in detail in the main text.
}
\label{fig:hiding}
\end{figure*}

\begin{figure*}[t!]
\center
\includegraphics[scale=0.75]{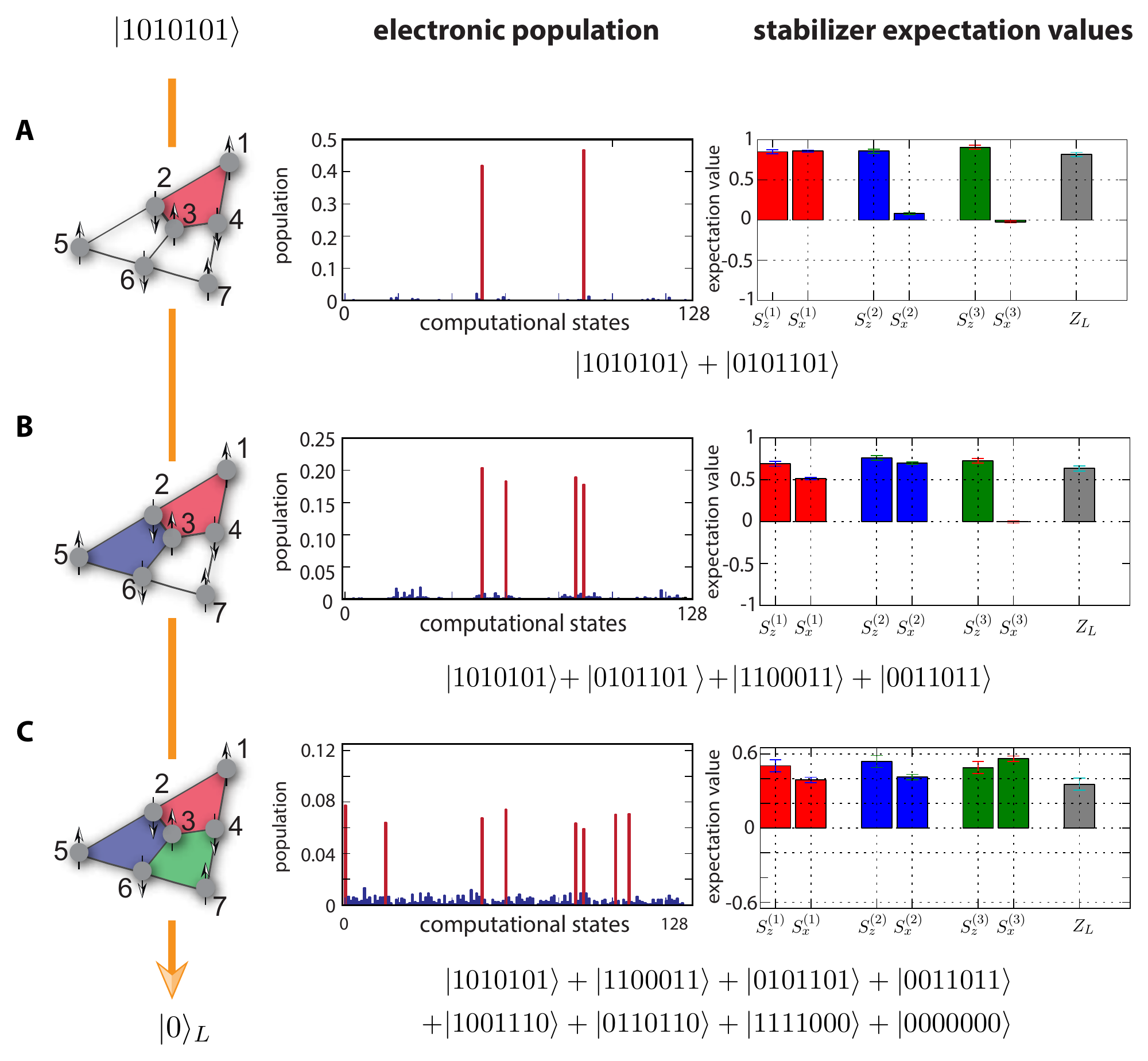}
\caption{\textbf{Initialization of the logical qubit (continued).} The step-wise encoding of the system in the logical state
$|0\rangle_{L}$ is observed by measurements of the $2^7 = 128$ electronic populations in the computational basis states, as well as by the measured pattern of
expectation values of the six stabilizers $\{S_{z}^{(1)},S_{z}^{(2)},S_{z}^{(3)},S_{x}^{(1)},S_{x}^{(2)},S_{x}^{(3)}\}$ together
with the logical stabilizer $Z_{L}$. The initial product state $|1010101\rangle$ is ideally a +1 eigenstate of all three $S_{z}^{(i)}$ stabilizers and of $Z_L$. \textbf{(A)} In the first step, GHZ-type entanglement is created between the four qubits belonging to the first (red) plaquette, which is signaled by the appearance of a non-vanishing, positive-valued $S_{x}^{(1)}$ expectation value. In the subsequent entangling steps acting on the second (blue) \textbf{(B)} and third (green) \textbf{(C)} plaquette, the system populates dominantly the expected four and eight computational basis states, and the created coherences show up in the non-zero expectation values of $S_{x}^{(2)}$ and $S_{x}^{(3)}$.
}
\label{fig:encoding}
\end{figure*}

The code space hosting logical states $\ket{\psi_L}$ is fixed as the simultaneous eigenspace of eigenvalue +1 of the six $S_x^{(i)}$ and $S_z^{(i)}$ stabilizer operators. The first step of the initialization of the 7-qubit system in the logical $\ket{0}_L$ state consists of converting the state of the 7 qubits, initialized by optical pumping in the state $\ket{1111111}$, into the state $\ket{1010101}$. This state fulfills the $Z$-type stabilizer constraints, as it is a +1 eigenstate of three $S_z^{(i)}$-stabilizers as well as of the logical operator $Z_L$. In the remaining three steps, the subsets of four qubits belonging to each of the three plaquettes of the code, are sequentially entangled to also fulfill the stabilizer constraints imposed by the three operators $S_x^{(i)}$. This is achieved by spectroscopically decoupling the ions hosting inactive qubits (e.g.~qubits 5, 6, and 7) not participating in the plaquette-wise entangling operation, as explained in Sec.~\ref{sec:SI_decoupling} and shown in Fig.~\ref{fig:hiding}. The MS gate operation, applied to the remaining active qubits (say, qubits 1, 2, 3, and 4) creates the four-qubit GHZ-type entanglement. Under this operation, for instance in the first step the state $|1010101\rangle$ is mapped onto the superposition $|1010101\rangle \pm i |0101101\rangle$ (see e.g.~\cite{s-mueller-njp-13-085007}). The $\pm \pi/2$ phase shift can be compensated for at any later instance during the remaining encoding sequence. To this end, at the end of the encoding sequence we apply an ac Stark shift compensation pulse $U_{Z}^{(1)}(\pm\pi/2)$ on qubit 1 located at the corner of the first (red) plaquette, which does not participate in entangling operations of the other two plaquettes. Thus, the four-qubit entangling operation in combination with the phase compensation pulse creates the $|1010101\rangle + |0101101\rangle$ superposition state after the first step of the encoding sequence. The successful preparation of this intermediate state can be seen from the measurement of electronic populations in the $2^7= 128$ computational basis states and the stabilizer operators, as shown in Fig.~\ref{fig:encoding}A. Here, the data shows the two dominant expected populations, as well as the GHZ-type coherence as signaled by the non-vanishing expectation value of $S_x^{(1)}$. In the second and third step, similarly, entanglement between the qubits of the second (blue) and third (green) plaquette is created, which is reflected (see Fig.~\ref{fig:encoding}B and C) by the appearance of electronic populations in the four and eight expected computational basis states, respectively, as well as by the step-wise build-up of non-vanishing coherences $\langle S_x^{(2)} \rangle$ and $\langle S_x^{(3)} \rangle$. After the third step and the application of the three phase compensation pulses the encoding of the system in the logical state $\ket{0}_L = |1010101\rangle + |0101101\rangle +
|1100011\rangle+|0011011\rangle + |1001110\rangle + |0110110\rangle + |1111000\rangle+
|0000000\rangle$ is signaled by positive (ideally +1) values of all six stabilizer and the logical $Z_L$ operator.

The described three-step sequence realizing plaquette-wise entangling operations to prepare the system in the state $\ket{0}_L$ works in principle by starting in any of the eight components of state $\ket{0}_L$. The advantage of choosing the initial input state $|1010101\rangle$ (instead of e.g.~$|0000000\rangle$) is that for this initial state the 7-qubit state is during part of the encoding sequence in a decoherence-free subspace (DFS), in which the system is insensitive to global phase noise, as caused by magnetic field and laser fluctuations of the collective local rotations, which to leading order affect all ions in the same way. This type of noise constitutes one of the dominant noise sources in our setup \cite{s-schindler-njp-15-123012}. Whereas the ideal quantum state after the second state (as given explicitly in Fig.~\ref{fig:encoding}B) still resides entirely in a DFS, the final state still benefits from partial protection with respect to the described global noise. This partial phase noise protection, in combination with the addressing-error-corrected decoupling and recoupling pulses as described in Sec.~\ref{sec:SI_decoupling} are essential to achieve the proper initialization of the system in the code space, despite the complexity and overall length of the complete encoding sequence (see Table~\ref{tab:Resources} for a detailed list and exact numbers of applied pulses).

\begin{table}
\begin{tabular}{|c|| c | c | c | c | c | c | c |}
\hline  & \hspace{7mm} & \hspace{7mm} & \hspace{7mm} & \hspace{7mm} & \hspace{7mm} & \hspace{7mm} & \hspace{7mm}  \\
\, qubit \, &  1 & 2 & 3 & 4 & 5 & 6 & 7 \\
& & & & & & &  \\ \hline
& & & & & & &  \\
\, pulses \, &  19 & 21 & 0 & 21 & 25 & 15 & 7   \, \\
& & & & & & &  \\ \hline
& & & & & & &  \\
\, ion \, &  7 & 2 & 4 & 6 & 1 & 3 & 5  \\
& & & & & & &  \\ \hline
\end{tabular}
\caption{\textbf{Encoding of physical qubits along the string of seven ions}. The table shows the number of addressed laser pulses each physical qubit is exposed to in the course of the encoding sequence, as shown in Fig.~\ref{fig:hiding}A. In order to minimize the effect of residual errors due to cross-talk with neighboring ions during spectroscopic decoupling and recoupling pulses, we distribute the physical qubits along the chain in such a way that the physical qubits exposed to many (few) addressed laser pulses are hosted by ions at the edge (center) of the chain.}
 \label{tab:qubit_to_ion_translation_table}
 \end{table}

\section{Quantitative characterization of the encoded logical states}
\label{sec:SI_state_characterization}

\subsection{Quantum state fidelities and overlap with the code space}
\label{sec:SI_fidelity}
The experimentally generated states are fully characterized by 7-qubit density matrices $\rho$, which could be reconstructed from 7-qubit quantum state tomographies \cite{s-nielsen-book}. However, as the ideal (target) states belong to the class of stabilizer quantum states \cite{s-gottesman-pra-54-1862}, the relevant figures of merit of the experimental states such as (i) the overall quantum state fidelities with the ideal states, (ii) the overlap with the code space, as well as (iii) the quantum state fidelities \textit{within} the code space can be exactly determined more efficiently from a reduced set of measurements.

For $n$ qubits, associated to a Hilbert space of dimension $d = 2^n$, a general state $\rho$ of the system can be expanded in the operator basis formed by all possible Pauli operators $W_k$, $k = 1, \ldots,d^2 = 4^n$, i.e.~the $n$-fold tensor product of the Pauli matrices $1, X, Y, Z$ for each qubit, with $\text{tr} (W_i W_j / d) = \delta_{ij}$. For a general quantum state, $\rho = \frac{1}{d} \sum_k \text{tr}(W_k\rho)\,W_k$, all $4^n$ expansion coefficients can contribute to the sum. For stabilizer states $\rho = \ket{\psi}\bra{\psi}$, for which $W_k \ket{\psi} = \pm \ket{\psi}$ if $W_k$ belongs to the the stabilizer group of state $\ket{\psi}$, only the $2^n$ coefficients corresponding to the set of stabilizer elements $W_k$ are non-zero: $\text{tr}(W_k\rho) = \pm 1$.

The quantum state fidelity of the experimental state $\rho$ with the ideal target stabilizer state as $\rho_t$ is given by
\begin{equation}
\label{eq:fidelity}
\mathcal{F}(\rho, \rho_t) = \text{tr}(\rho_t \, \rho) = \frac{1}{d} \sum_k \text{tr}(W_k\rho_t) \text{tr}(W_k\rho).
\end{equation}
Since there are $2^n$ non-zero coefficients $\text{tr}(W_k\rho_t)$ contributing to this sum, it is sufficient to measure the expectation values of the corresponding $2^n$ Pauli operators $W_k$, $\langle W_k \rangle_\rho = \text{tr}(W_k\rho)$, to exactly determine the quantum state fidelity according to Eq.~(\ref{eq:fidelity}). For the present case of encoded quantum states given by 7-qubit stabilizer states, this requires the measurement of only 128 expectation values $\langle W_k \rangle_\rho$, without the need to reconstruct the full density matrix $\rho$.

Note that the six ideal encoded states can be written as a product of two projectors
\begin{equation}
\rho_t = P_{\pm O_L} P_{CS}.
\end{equation}
Here, $P_{CS}$ denotes the projector onto the code space, defined as the simultaneously +1 eigenspace of the six stabilizers (generators) $S_x^{(i)}$ and $S_z^{(i)}$, $i=1,2,3$,
\begin{equation}
\label{eq:P_CS}
P_{CS} = \prod_{i=1}^{3} \frac{1}{2} (1+ S_x^{(i)}) \prod_{j=1}^{3} \frac{1}{2} (1+ S_z^{(j)}),
\end{equation}
and $P_{\pm O_L} = \frac{1}{2} (1 \pm O_L)$ is the projector onto the desired logical state within the code space, concretely:
\begin{align}
\label{eq:P1_projector}
\rho_{\ket{1}_L} &= \ket{1} \bra{1}_L = \frac{1}{2}(1 -  Z_L) P_{CS}, \\
\label{eq:P0_projector}
\rho_{\ket{0}_L} &= \ket{0} \bra{0}_L = \frac{1}{2}(1 + Z_L) P_{CS}, \\
\rho_{\ket{\pm_x}_L} &= \ket{\pm_x} \bra{\pm_x}_L = \frac{1}{2}(1 \pm X_L) P_{CS}, \\
\label{eq:PY_projector}
\rho_{\ket{\pm_y}_L} &= \ket{\pm_y} \bra{\pm_y}_L = \frac{1}{2}(1 \pm Y_L) P_{CS}.
\end{align}
Note that the logical operator $Y_L = + i X_L Z_L = - \prod_{i=1}^{7} Y_i$ for the 7-qubit color code\vspace{2mm}.

\textbf{The quantum state fidelity} of the experimentally generated state $\rho$ with the ideal target state, say $\rho_{\ket{1}_L}$, is given by $\mathcal{F}(\rho, \rho_{\ket{1}_L}) = \frac{1}{128} \sum_{k=1}^{128} \langle W_k \rangle$, which is the equal-weighted sum of expectation values of the 128 Pauli operators appearing in the product of projectors in Eq.~(\ref{eq:P1_projector}), $W_1 = 1$, $W_2 = - Z_L$, $\ldots$, $W_{128} = - Z_L S_x^{(1)}S_x^{(2)}S_x^{(3)} S_z^{(1)}S_z^{(2)}S_z^{(3)} = - X_1 Z_2 X_3 Z_4 X_5 Z_6 X_7$. Similarly, there are a set of 128 operators to be measured for the characterization of the other logical states, according to the combinations appearing in Eqs.~(\ref{eq:P0_projector}) - (\ref{eq:PY_projector})\vspace{2mm}.

\textbf{The overlap with or population in the code space} is given by $p_{CS} = \text{tr}(P_{CS} \rho) = \frac{1}{64} \sum_{k=1}^{64} \langle W_k \rangle$, where the sum extends over the expectation values of the 64 Pauli operators contained in the expansion of the projector onto the code space $P_{CS}$ (see Eq.~(\ref{eq:P_CS})).

The projected (and normalized) density matrix of the experimental state $\rho$ within the code space is given by $\rho_{CS} := P_{CS} \rho P_{CS} / p_{CS}$. Thus, for a given experimental state $\rho$, the \textbf{quantum state fidelity within the code space} with the ideal target state $\rho_t = P_{\pm O_L} P_{CS}$ is given by
\begin{align}
\mathcal{F}(\rho_{CS}, \rho_t) & = \text{tr}(P_{\pm O_L} P_{CS} \, P_{CS} \rho P_{CS}) / p_{CS} \nonumber \\
& =  \text{tr}(P_{\pm O_L} P_{CS} \rho) / p_{CS} \nonumber \\
& = \mathcal{F}(\rho, \rho_t) / p_{CS}
\end{align}
From the latter expression one sees that the overlap fidelity $\mathcal{F}(\rho, \rho_t) = p_{CS} \mathcal{F}(\rho_{CS}, \rho_t)$ is indeed given by the combination of the overlap with the code space and the fidelity with the target state \textit{within} the code space.

\begin{figure}[t!]
\center
\includegraphics[width=0.7\columnwidth]{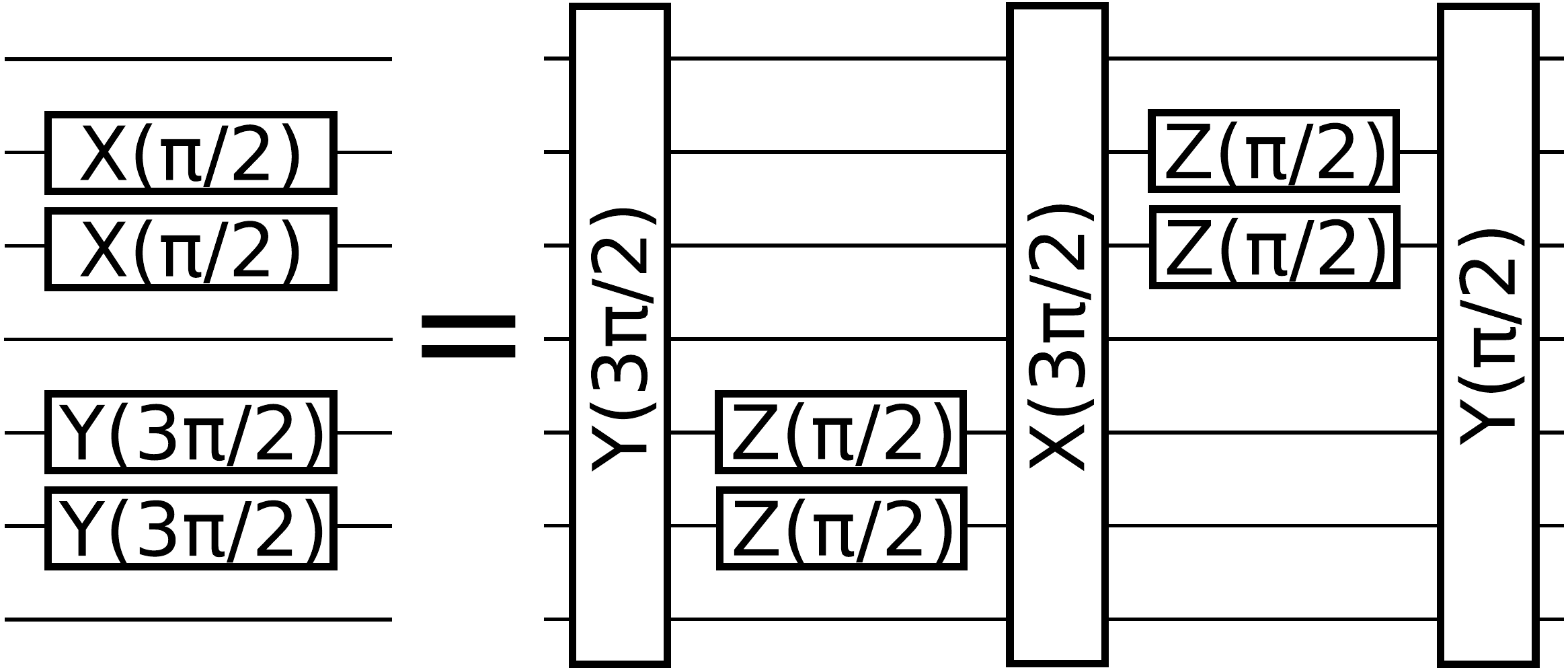}
\caption{Experimental pulse sequence for the measurement of the Pauli operator $S_z^{(1)} S_x^{(2)} =  - Z_1 Y_2 Y_3 Z_4 X_5 X_6$, which is one of 128 operators that are measured for the quantitative characterization of encoded logical states.}
\label{fig:tomo_pulses}
\end{figure}

Measurement of the required set of 128 operators $\{W_k\}$ for a given encoded logical target state $\ket{\psi_L}$, requires the application of a sequence of local unitaries after the preparation sequence for $\ket{\psi_L}$. As in standard quantum state tomography \cite{s-nielsen-book}, this is needed to transform a given Pauli operator $W_k$ into the standard measurement basis ($Z$) in which the fluorescence measurements are physically realized. Pulse sequences for these basis transformations are determined by writing the local unitaries as a product of the available set of operations (as described in Sec.~\ref{sec:SI_experimental_system}), and subsequently determining the required rotation parameters analytically. As an example, Fig.~\ref{fig:tomo_pulses} displays the pulse sequence used for the measurement of the Pauli operator $S_z^{(1)} S_x^{(2)} =  - Z_1 Y_2 Y_3 Z_4 X_5 X_6$.

The error bars of the measured quantum state fidelities were determined by a Monte-Carlo method,
resampling the measured fluorescence data based on the uncertainty given by the limited number of measurement
cycles. Therefore, the measurement data was randomly resampled using a multinomial
distribution with a statistical uncertainty of $\sqrt{p(1-p)/N}$ for each measured
probability $p$ and number of repetitions $N$. The fidelity $\mathcal{F}(\rho, \rho_t)$,
$\mathcal{F}(\rho_{CS}, \rho_t)$ and the populations in the code space $p_{CS}$ are calculated for
from the mean value and standard deviation of all Monte-Carlo results.

\subsection{Entanglement properties of the encoded logical qubit}
\label{sec:SI_entanglement_properties}

We study the entanglement properties of the encoded 7-qubit states using the method of entanglement witnesses. Following Ref.~\cite{s-bourenane-prl-92-087902}, we construct witness operators $\mathcal{W}$, which are able to detect entanglement between different bi-partitions of the 7-qubit system, provided that the experimentally generated states have a sufficiently high quantum state fidelity with the ideal target states. From the explicit form of the ideal logical state $\ket{1}_L = |0101010\rangle + |1010010\rangle + |0011100\rangle + |1100100\rangle + |0110001\rangle + |1001001\rangle + |0000111\rangle+ |1111111\rangle$ it can be verified that with respect to any bi-partition of the 7 qubits into a part A that consists of two qubits (say, e.g. qubits 1 and 2) and a part B consisting of the complementary five qubits (say qubits 3-7), the square of the largest Schmidt coefficient of the state is 1/4. Thus, we consider the witness operator $\mathcal{W} = \frac{1}{4} \mathbf{1} - \ket{1} \bra{1}_L$. It signals entanglement of the two parts A and B if the expectation value $\langle \mathcal{W} \rangle$ in the experimental state $\rho$ is negative. Since $\langle \mathcal{W} \rangle_\rho = 1/4 - \mathcal{F}(\rho, \rho_{\ket{1}_L}$), this is the case for quantum state fidelities larger than 25\%, in which case the experimental state is not separable with respect to any bipartition of 2 and 5 qubits. This implies the presence of entanglement of at least six qubits, which in turn indicates the mutual entanglement of all three plaquettes, as there is no combination of six qubits which only involves two plaquettes. Similarly, a quantum state fidelity threshold of 25\% holds for the other five logical states (see Eqs.~(\ref{eq:P0_projector}) - (\ref{eq:PY_projector})). These states are locally (i.e.~up to single-qubit rotations) equivalent to $\rho_{\ket{1}_L}$ -- which can be seen, e.g.,~directly from the transversal character of the Clifford gate operations by which they can be transformed into one another.

For the experimentally generated encoded states \{$\ket{0}_L$, $\ket{1}_L$ and $\ket{+_x}_L$\} the measured quantum state fidelities with the ideal target states of \{32.7(8), 28(1), 33(1)\}\% surpass the threshold value of 25\% by more than \{9, 3, 8\} standard deviations.

\section{Absence of local order and presence of global order in the color code state}
\label{sec:SI_topological_order}

\textbf{General remarks and background:} One of the intriguing features of topological quantum phases of matter is that they evade a conventional characterization by local order parameters, such as the local magnetization or (anti-)ferromagnetic correlations of neighboring spins in magnetic materials \cite{s-wen-book}. In contrast, the detection of quantum order and the distinction of different quantum phases in topological quantum systems can be achieved by means of global observables which are able to retrieve correlations which extend over the entire many-particle system. In topological quantum codes, such as Kitaev's toric code \cite{s-kitaev-annalsphys-303-2} and related models \cite{s-raussendorf-njp-9-199}, as well as the in topological color codes \cite{s-bombin-prl-97-180501, s-bombin-prl-98-160502, s-bombin-prb-75-075103}, global quantum order can be detected and manipulated by acting collectively on groups of physical qubits, which are located along strings that extend over the whole spin system \cite{s-wen-book, s-kitaev-annalsphys-303-2}.

\begin{figure*}[t!]
\center
\includegraphics[scale=1]{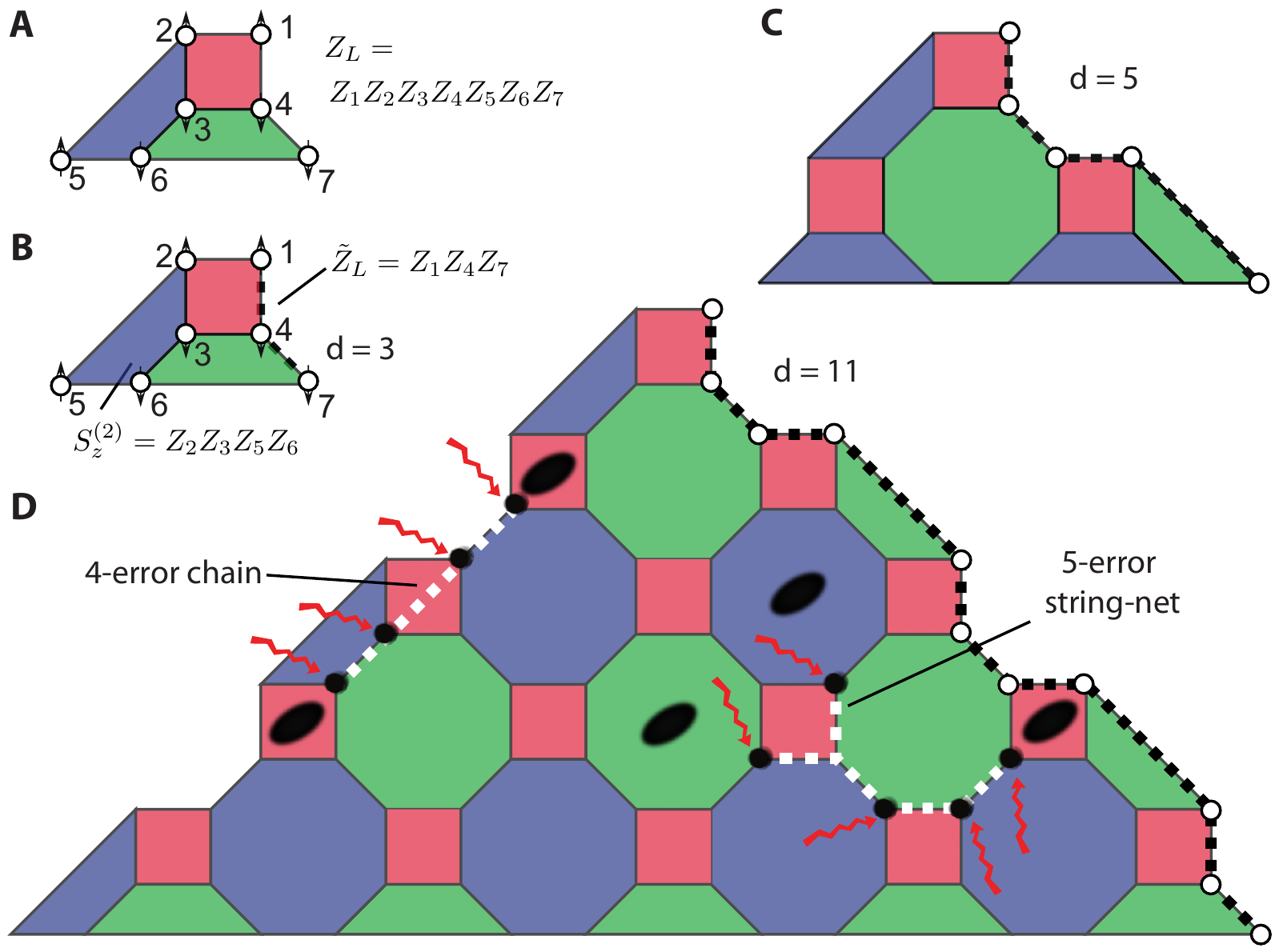}
\caption{\textbf{2D color codes constructed on increasingly larger lattices, each hosting one logical qubit}. The displayed examples belong to the class of 2D color codes defined on triangular 4-8-8 lattices - these lattices are three-colorable and each vertex (except those which are in involved in plaquettes at the boundary) is surrounded by three plaquettes of different color, with 4, 8, and 8 physical qubits per plaquette, respectively. 2D color codes on 4-8-8 lattices are of particular interest, as they allow one to implement the $K$-gate in a transversal way, and as a consequence, the whole Clifford group of logical gate operations \cite{s-bombin-prl-97-180501}. \textbf{(A)} The smallest lattice involving 7 physical qubits (indicated by white circles) corresponds to the 7-qubit color code implemented in this work, consisting of three plaquettes of different color, each associated with a four-qubit $S_x^{(i)}$ and $S_z^{(i)}$ stabilizer. The logical operator $Z_L= Z_1 Z_2 Z_3 Z_4 Z_5 Z_6 Z_7$ acts on all 7 physical qubits of the code layer. Note that in the present visual representation the plaquettes are distorted as compared to the more symmetric structure in Fig.~1A of the main text, however, the quantum code is obviously unchanged. The 7-qubit code structure constitutes the minimal triangular 2D code structure, from which codes on larger lattices with boundaries \cite{s-bravyi-arXiv:quant-ph/9811052} can be constructed (see Fig.~\ref{fig:lattices}C and D). \textbf{(B)} Same code structure as in Fig~\ref{fig:lattices}A. The action of global logical operator $Z_L$ in the code space is equivalent to the action of the string operator $\tilde{Z}_L =  Z_1 Z_4 Z_7$ acting on the three qubits (shown as white dots, connected by a black dashed line) along the right boundary of the triangular lattice. The string operator $Z_L$ is obtained from multiplication of $Z_L$ by an element from the stabilizer group, here the $S_z^{(2)}$ stabilizer attached to the blue plaquette. The 7-qubit color code has a logical distance $d=3$, which implies that the code can correct $(d-1)/2 = 1$ error on any of the physical qubits \cite{s-nielsen-book}. Figure \textbf{(C)} shows a distance $d=5$ color code that can correct up to 2 arbitrary errors that can occur on any of the 17 physical qubits of the code. For simplicity only the 5 physical qubits (white dots), which participate in the shown logical operator $\tilde{Z}_L$ (connected by the dashed line), are shown. Here, the logical $\tilde{Z}$ string operator involving 5 physical qubits can be obtained by multiplication of the $Z_L$ operator, which acts on all physical qubits, with the $Z$-type stabilizers of all blue plaquettes. \textbf{(D)} The figure shows a distance $d=11$ color code, able to correct $5$ errors. The dashed line connects 11 qubits participating in one possible representation of the logical $\tilde{Z}_L$ string operator. In the left part of the lattice, a scenario is shown where, a sequence of single-qubit errors of the same type, e.g.~bit-flip errors (indicated by red wiggled arrows), affects four physical qubits (marked by black filled circles). This physical error chain results in an error chain (white dashed line), where the (equally-colored) plaquettes at the two end points of the chain are in -1 eigenstates of the corresponding  ($S_z^{(i)}$) stabilizer operators - these stabilizer violations are indicated as black-shaded ellipses on the red plaquettes. In color codes, sequences of physical errors (such as the combination of 5 physical errors shown in the right part of the lattice) can not only be connected by chains, but also result in error string-nets (dashed white line) that undergo branching \cite{s-bombin-prl-97-180501, s-bombin-prl-98-160502, s-bombin-prb-75-075103, s-wen-book}. In the displayed example, the string-net has three endpoints terminating at the three plaquettes of different colors, which will display stabilizer violations (-1 eigenvalues) of the associated plaquette operators in the syndrome measurement.}
\label{fig:lattices}
\end{figure*}

Figure \ref{fig:lattices} shows examples of 2D triangular color codes of increasing lattice size. The stabilizer operators act locally, i.e.~on physical qubits that are involved in the corresponding plaquettes, as opposed to global logical operators of the quantum code, such as the logical $Z_L$ operator which acts bitwise on all qubits of the 2D layer. Note that the code space is defined as the simultaneous +1 eigenspace of all plaquette stabilizers, thus $S_i \ket{\psi_L} = + \ket{\psi_L}$ for any encoded quantum state $\ket{\psi_L}$. This property allows one to transform the logical operator $Z_L$ into logically equivalent operators $\tilde{Z}_L$ by multiplication with (combinations of) stabilizers: For instance, for the smallest color code (see Fig.~\ref{fig:lattices}A), involving seven physical qubits (see Fig.~\ref{fig:lattices}A), $Z_L= Z_1 Z_2 Z_3 Z_4 Z_5 Z_6 Z_7$ and $Z_L \ket{\psi_L} = Z_L S_z^{(2)}\ket{\psi_L} = Z_L Z_2 Z_3 Z_5 Z_6 \ket{\psi_L} = Z_1 Z_4 Z_7 \ket{\psi_L}$. This shows that the string operator $\tilde{Z}_L =  Z_1 Z_4 Z_7$ (acting on the qubits connected by the black dashed line or string shown in Fig.~\ref{fig:lattices}B) in the code space is fully equivalent to $Z_L$. Whereas the logical operator $\tilde{Z}_L$ only acts on three instead of all seven physical qubits, it is still a global operator as it extends over the entire side length of the triangular code (see Fig.~\ref{fig:lattices}B). In larger 2D color codes, such as the examples shown in Figs.~\ref{fig:lattices}C and D, the corresponding logical global operators involve more physical qubits, here 5-qubit and 11-qubit string operators, respectively.

Topological color codes display Abelian topological order: for the purpose of quantum error correction stabilizer violations (-1 eigenstates) associated to plaquettes signal the occurrance of one or several errors. However, from a condensed matter perspective the code space of the topological code can be interpreted as the ground state manifold of a many-body Hamiltonian, whose lowest-energy states are characterized by the fact that all stabilizer +1 constraints are simultaneously fulfilled \cite{s-bombin-prl-97-180501, s-bombin-prb-75-075103}. Here, stabilizer violations correspond to quasi-particle excitations, localized on plaquettes. These quasi-particle excitations show unusual particle exchange statistics: when quasi-particles are winded around each other (braiding) the system returns its initial quantum state, which however differs from the initial state by a phase factor which neither corresponds to the one of bosons nor fermions -- thus termed Abelian anyons \cite{s-wilczek-prl-49-957, s-leinass-nuocimb-37-1, s-kitaev-annalsphys-303-2}. The topological order within the code space or ground state manifold is intimately related to the statistics and the dynamical behavior of stabilizer violations or quasi-particle excitations \cite{s-wen-book}. In color codes, stabilizer violations can be connected by error chains with two end points terminating on plaquettes with eigenvalue -1 (see Fig.~\ref{fig:lattices}D and figure caption for details). Alternatively, error chains can undergo branching and form error string nets (see Fig.~\ref{fig:lattices}D for an example). This rich dynamical behavior of quasi-particle excitations (stabilizer violations) is a consequence of the $\mathbb{Z}_2 \times \mathbb{Z}_2 $ Abelian topological order in color codes \cite{s-bombin-prl-97-180501, s-bombin-prb-75-075103}. This is in contrast to Kitaev's toric code \cite{s-kitaev-annalsphys-303-2}, where stabilizer violations always appear in pairs and are connected exclusively by linear error chains, reflecting the different type ($\mathbb{Z}_2$) of Abelian topological order in this model.

\begin{figure*}[t!]
\center
\includegraphics[scale=0.7]{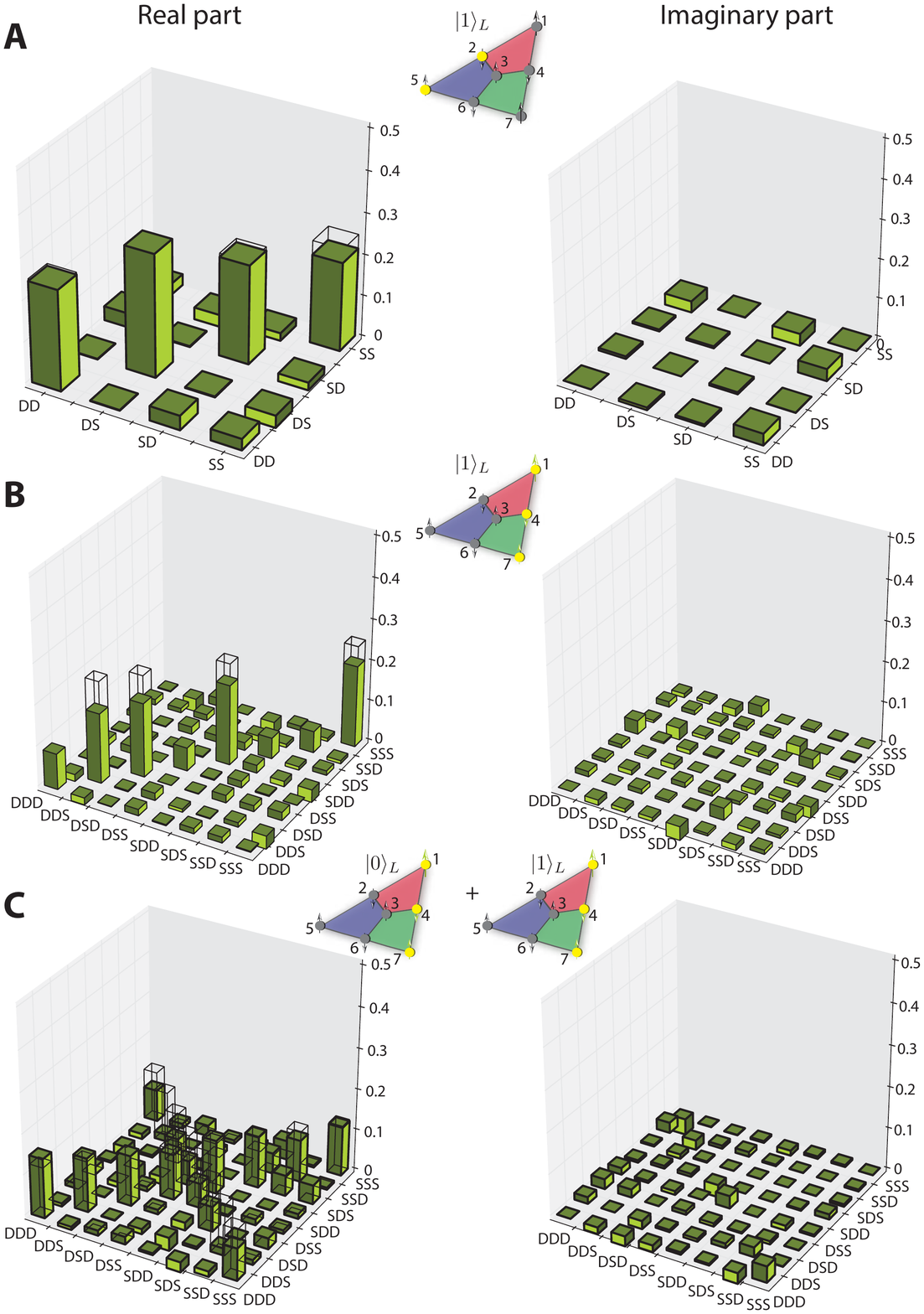}
\caption{\textbf{Absence of local and presence of global quantum order
    in the topologically encoded qubit.} Real and imaginary elements
  of reconstructed 2- and 3-qubit density matrices. Matrix elements of
  the ideal states are indicated as transparent bars. The respective
  encoded logical state is schematically indicated for each
  sub-figure, together with the subsets of qubits (marked as yellow
  filled circles) on which quantum state tomographies have been
  performed. Electronic populations $D$ ($S$) correspond populations
  in the computational $\ket{0}$ ($\ket{1}$) state. \textbf{(A)}
  Measured 2-qubit density matrix of qubits 2 and 5, clearly
  indicating the large overlap (98.3(2)\%) with the ideal completely
  mixed 2-qubit density matrix (incoherent equal-weighted mixture of
  the four computational basis states). \textbf{(B)} The displayed
  reduced 3-qubit density matrix (of qubits 1, 4 and 7) has a quantum
  state fidelity of 85(2)\% with the ideal state, which is an
  incoherent equal-weighted mixture of the four 3-qubit basis states
  $\ket{111}$, $\ket{001}$, $\ket{010}$ and $\ket{100}$. The measured
  reduced 3-qubit density matrix (of qubits 1, 4 and 7) for the
  logical qubit initially prepared in $\ket{+_x}_L$, yields a quantum
  state fidelity of 83(2)\% with the ideal state.}
\label{fig:suppl_tomo}
\end{figure*}

\vspace{1mm}\textbf{Experimental study of the topological order of the 7-qubit color code}: For the implemented 7-qubit code, local operators thus refer to one or two-qubit operators, whereas global operators such as $\tilde{Z}_L$ at least involve 3 physical qubits. For the experimental study of the quantum order of the encoded qubit, we first prepare the 7-qubit system in the logical state $\ket{1}_L$ and $\ket{+_x}_L$, and subsequently perform a series of quantum state tomography measurements on subsets of two and three qubits:

\begin{itemize}
\item
We perform 2-qubit tomographies on all 21 subsets of two out of seven ions, and reconstruct the reduced two-qubit density matrices. We find that these yield an average Uhlmann-fidelity \cite{s-uhlmann-rep-math-phys-9-273} with the two-qubit completely-mixed state of 98.3(2)\% (the largest and smallest obtained fidelity values are 99.0(4)\% and 97.7(8)\%, respectively). Figure \ref{fig:suppl_tomo}A shows as a representative example the elements of the reconstructed reduced density matrix of qubits 2 and 5.
\item
From a 3-qubit state tomography on qubits 1, 4 and 7, we determine the expectation value of the global string operator $\tilde{Z}_L =  Z_1 Z_4 Z_7$, which yields non-vanishing  3-qubit correlations $\langle Z_1 Z_4 Z_7 \rangle = - 0.46(6)$, clearly signalling the presence of global quantum order. Figure \ref{fig:suppl_tomo}B shows the reconstructed reduced 3-qubit density matrix.
\item
Furthermore, we also prepared the logical superposition state $\ket{+_x}_L = (\ket{0}_L + \ket{1}_L) /\sqrt{2}$. Here the global order becomes manifest in non-vanishing three-qubit correlations of the $X$-type string operator $\tilde{X}_L = X_1 X_4 X_7$, which is equivalent to the logical $X_L = X_1 X_2 X_3 X_4  X_5 X_6  X_7$ operator ($\tilde{X}_L = X_L S_x^{(2)}$). We find $\langle X_1 X_4 X_7 \rangle$ = 0.40(5). See Fig.~\ref{fig:suppl_tomo}C for the reconstructed reduced 3-qubit density matrix.
\end{itemize}
These measurements confirm the topological character of the encoding of the logical qubit, as they clearly demonstrate the absence of local order in the experimental state, as well as the presence of global quantum order, which for the present size becomes manifest in non-vanishing 3-qubit correlations.

\section{Quantum error detection and complete experimental syndrome table}
\label{sec:SI_syndrome}
\textbf{General remarks and background:} In the ``traditional" approach towards FTQC \cite{s-shor96a, s-knill-science-279-342, s-aharonov97, s-gottesman-pra-57-127, s-aliferis_etal06, s-preskill97}, elementary quantum codes, such as e.g.~the 5-qubit (non-CSS-type) code \cite{s-laflamme-prl-77-198, s-bennett-pra-54-3824}, the 7-qubit code, first suggested by Steane \cite{s-steane-prl-77-793} and implemented in this experiment, or the 9-qubit code proposed by Shor \cite{s-shor-pra-52-2493(R)}, are concatenated \cite{s-knill-arXiv:quant-ph/9608012}. Here, protection of encoded logical quantum information is achieved by encoding information redundantly in several physical qubits, each of which is again encoded in a new layer of physical qubits, and so forth. This requires a considerable overhead in qubits, which grows exponentially in the number of encoding layers used in the encoding hierarchy. Furthermore error detection and correction quantum circuitry generally involves operations between far-distant qubits of the code.

In topological quantum error correcting codes, protection of errors is achieved in a qualitatively different way: logical qubits are encoded in large lattice systems of physical qubits. Here, due to the topological nature of the codes, quantum error correction operations, such as measurements of the stabilizer or check operators defining the code space respect the spatial structure of the underlying lattice, i.e.~they act on groups of physically neighboring qubits belonging for instance to the plaquettes of the lattice. Due to this locality property topological codes are ideally suited to be embedded in physical 2D architectures \cite{s-ladd-nature-464-45}, where individual qubits and groups of few adjacent qubits can be addressed and manipulated locally \cite{s-kitaev-annalsphys-303-2, s-bravyi-arXiv:quant-ph/9811052, s-bravyi-j-math-phys, s-bennett-pra-54-3824, s-bombin-prl-97-180501, s-bombin-prl-98-160502}. Robustness of logical qubits is achieved by encoding logical qubits in larger and larger lattice structures, see Fig.~\ref{fig:lattices} for examples. Roughly speaking, such large code structures can tolerate more and more errors occurring locally on physical qubits, before the logical quantum information, encoded in global (topological) properties of the many-qubit system, cannot be recovered after the occurence of too many errors on the register. Provided that single-qubit errors occur with a low enough error rate, and the error syndrome formed by the stabilizer information can be measured with sufficient accuracy, quantum information stored in large topological codes can be protected from errors and processed fault-tolerantly \cite{s-dennis-j-mat-phys-43-4452}. The robustness and error thresholds have been studied analytically and numerically for various topological quantum codes, including color codes \cite{s-katzgraber-prl-103-090501, s-katzgraber-pra-81-012319, s-wang-quant-inf-comp-10-780, s-landahl-arXiv:1108.5738, s-ohzeki-pre-80-011141, s-andrist-njp-13-083006, s-bombin-prx-2-021004, s-jahromi-prb-87-094413},  and Kitaev's toric code and related models \cite{s-dennis-j-mat-phys-43-4452, s-wang-ann-phys-303-31, s-ohno-nucl-phys-b-697-462, s-takeda-jphysa-38-3751, s-raussendorf-prl-98-190504, s-raussendorf-njp-9-199, s-fowler-arXiv:1209.0510, s-al-shimary-njp-15-025027, s-brell-arXiv:1311.0019}. Here, threshold vaues are code-dependent and also strongly depend on the noise model considered, typically yielding error-per-operation thresholds on the order of $10^{-2}$ to $10^{-1}$ for phenomenological noise models and somewhat lower threshold values on the order of $10^{-3}$ to $10^{-2}$ for circuit noise models, where errors are taken into account at the level of imperfections in the quantum circuitry which is required for the readout of the error syndrome. The deduction of logical error classes requires classical processing of the measured syndrome information, a task for which decoding algorithms have been developed \cite{s-edmonds-oper-research-1965, s-dennis-j-mat-phys-43-4452, s-duclos-cianci-prl-104-050504, s-wang-quant-inf-comp-10-780, s-sarvepalli-pra-85-022317, s-stephens-arXiv:1402.3037}.

\vspace{1mm}\textbf{Experimental error detection for the 7-qubit code}:
The implemented 7-qubit code, shown in Fig.~1A of the main text and in Fig.~\ref{fig:lattices}A, is a CSS code, and has a logical distance $d=3$, which implies that it can correct $(d-1)/2 = 1$ physical error, regardless on which physical qubit of the code the error occurs \cite{s-steane-prl-77-793}. This logical distance of the topological quantum code is directly related to the geometric size of the code, namely it corresponds to the side length of the triangular code structure (see Fig.~\ref{fig:lattices}B). Larger 2D lattices, such as the two examples displayed in Figs.~\ref{fig:lattices}C and D, can host a logical qubit of bigger logical distance, which allow one to correct a larger number $n$ of single-qubit errors ($d=5$ and $n=2$ in (C), and $d=11$ and $n=5$ in (D)).

To study the error detection capability of the experimental 7-qubit code, we induce all 21 single-qubit errors on the logical qubit initially prepared in the state $\ket{0}_L$, and record the induced error syndromes. The errors are induced coherently by the corresponding Pauli operator $X_i$, $Y_i$ or $Z_i$. The $Z_i$ errors can be induced directly by a single AC Stark shift operation using the addressed beam (see Sec.~\ref{sec:SI_coherent_gates}), whereas the $X_i$ and $Y_i$ are realized by a combination of the collective local operations and single-ion $Z_i$ operations.

\begin{figure}[t!]
\center
\includegraphics[scale=0.8]{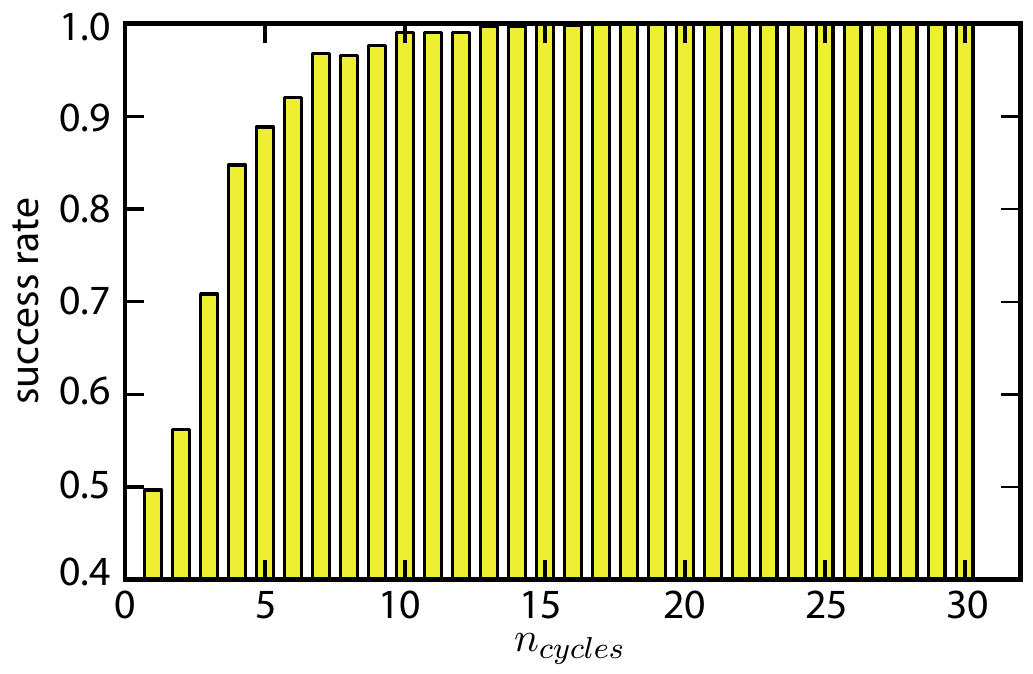}
\caption{Simulation of the success rate to identify the right error syndrome for a given number
of measurement cycles $n_{cycles}$. The success rate converges rapidly towards 100\% for $n_{cycles} > 20$ measurement cycles, implying that in these cases
the error syndromes can be clearly distinguished and perfectly associated to the induced physical single-qubit error. The number of Monte-Carlo samples used to determine each data point is 5000.}
\label{fig:MC}
\end{figure}

\begin{figure*}[t!]
\center
\includegraphics[scale=0.7]{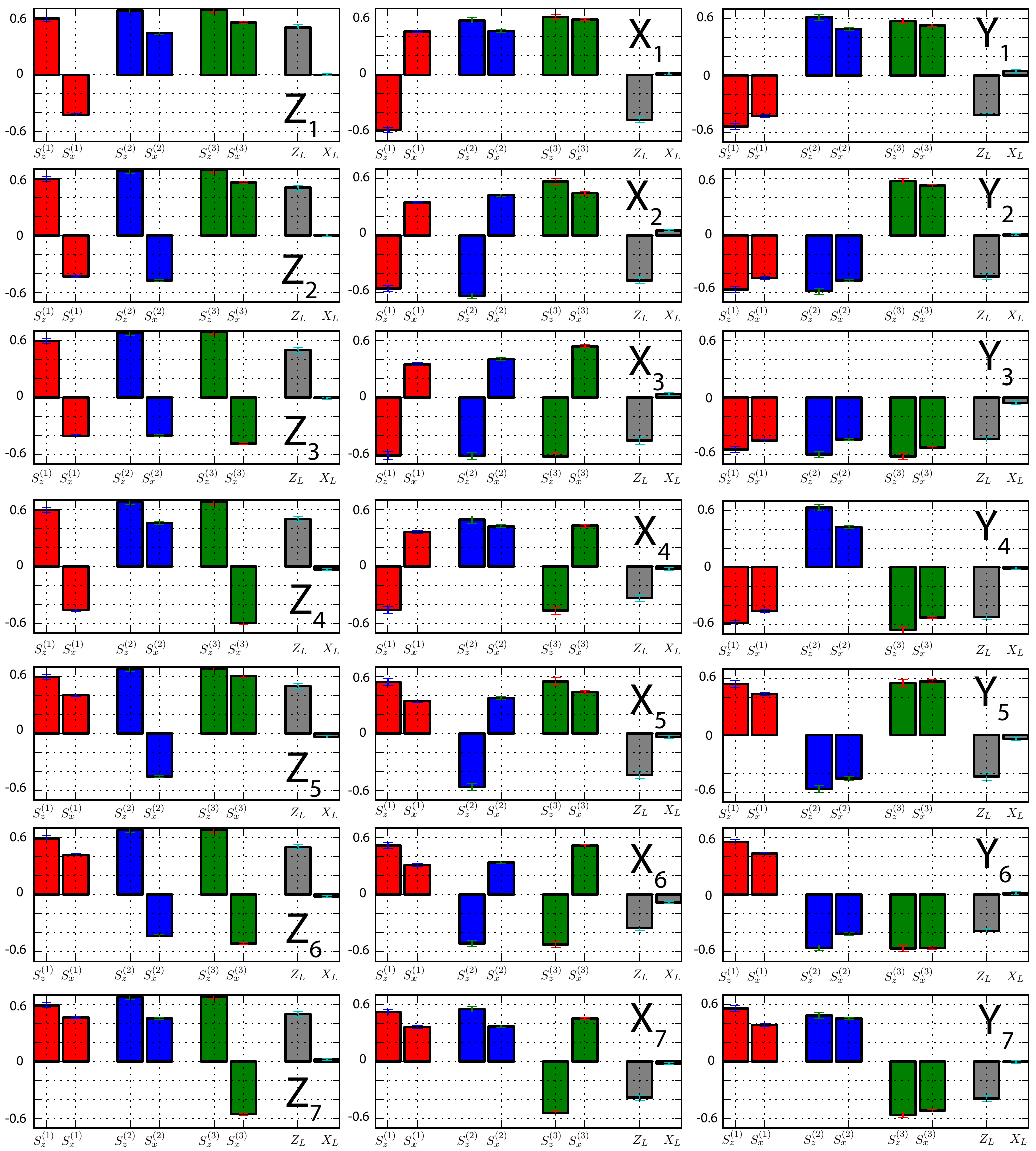}
\caption{Table of the recorded error syndromes for the logical qubit initially prepared in the encoded state $\ket{0}_L$, and subsequently exposed to all 21 single-qubit errors. Note that the error syndromes induced by a single qubit phase flip $Z$ on qubit 1 or qubit 4 are indistinguishable from the error syndromes caused by two $Z$ errors on qubits 2 and 5 (see Fig.~2E in the main text) or on qubits 3 and 5 (see Fig.~2F in the main text).}
\label{fig:syndrome}
\end{figure*}

Figure \ref{fig:syndrome} shows the complete syndrome table. Within our experimental uncertainties the measured error syndromes can unambiguously associated with the error syndromes induced by single-qubit errors. To quantify the classification quality of the individual measured syndromes, we perform a Monte-Carlo based simulation of the measured fluorescence data. The idea is to sample the data set using a multinomial distribution and calculate for each sampled data the stabilizer pattern. For each of the simulated stabilizer patterns, the success of correctly assigning the observed error syndrome to the induced single-qubit error is quantified by calculating the classical trace distance between the sampled stabilizer distributions $S_{i}^{(sample)}$ and the 21 measured reference stabilizers $S_{i}^{(ref)}$ of Fig. \ref{fig:syndrome}. The trace distance $D$ between the two classical distributions of the six stabilizer expectation values is given by
\begin{equation*}
D=\sqrt{\sum_{i=1}^{6}\left[S_{i}^{(ref)}-S_{i}^{(sample)}\right]^{2}},
\end{equation*}
\noindent and yields $D=0$ if the distributions are equal. The pattern of stabilizers, as generated by the Monte-Carlo method, is then associated to the reference syndrome for which the trace distance is minimal. Figure~\ref{fig:MC} shows the success rate of assigning the right error syndrome
(e.g. for a $Y$-error on qubit 3) as a function of the number of measurement cycles $n_{cycles}$. The success rate is defined
by the fraction of cases, in which the error syndrome has been correctly assigned to the corresponding single-qubit reference error syndrome, divided by the total number of attempts. It can be seen clearly that the success rate converges rapidly to 100\% after about $n_{cycles}$ = 20 measurement cycles. In the experimental measurements of the set of $S_x^{(i)}$ and $S_z^{(i)}$ stabilizers used for the error syndrome, we used 1000 cycles.

Correction of two or more single-qubit errors is beyond the error correction capacity of the 7-qubit code, and requires the encoding of a logical qubit in more physical qubits (such as, e.g., the distance $d=5$ 2D color code shown in Fig.~\ref{fig:lattices}C). We experimentally study the failure of the 7-qubit code for two cases: Figures 2E and F of the main text show the recorded error syndromes after inducing two single-qubit phase flip errors $Z$ on qubits 5 and 2 (Fig.~2E), and qubits 5 and 3 (Fig.~2F), respectively. The comparison with the single-qubit syndrome table in Fig.~\ref{fig:syndrome} shows that the recorded error syndromes are indistinguishable from the error syndromes induced by a $Z_1$ error (first column, first error syndrome) and a $Z_4$ error (first column, fourth error syndrome), respectively. Consequently, the erroneous deduction that a $Z_1$ ($Z_4$) error has happened, instead of the physical $Z_5$ and $Z_2$ errors ($Z_5$ and $Z_3$ errors) effectively result in the application of the operators $Z_1 Z_2 Z_5$ ($Z_3 Z_4 Z_5$) to the encoded qubit. These operators are equivalent to the logical operator $Z_L = Z_1Z_2 Z_3 Z_4 Z_5 Z_6 Z_7$ (since $Z_1 Z_2 Z_5 = Z_L S_z^{(3)}$ and $Z_3 Z_4 Z_5 = Z_L S_z^{(1)} S_z^{(3)}$) and thus result in a logical phase flip error $Z_L$ on the encoded state.

\section{Encoded Clifford quantum gate operations on the logical qubit}
\label{sec:SI_Clifford_gates}

In topological color codes, quantum information is processed directly within the code space \cite{s-bombin-prl-97-180501, s-bombin-prl-98-160502} -- leaving the code space and the presence of stabilizer violations always indicate the occurrence of an error. This is different from schemes of topological quantum computing, which rely on the controlled generation and manipulation of quasi-particle excitations for the implementation of logical gate operations \cite{s-nayak-rmp-80-1083}. 2D color codes enable a transversal (i.e~bit-wise) implementation of the entire group of logical Clifford gate operations. Here, we implement the generating gate operations $Z_L$, $X_L$, $H_L$ and $K_L$ of the single-qubit Clifford group on the topologically encoded qubit.

\subsection{Implementation of logical single-qubit Clifford gate operations}
The logical operators defining the logical qubit, $Z_L = Z_1Z_2 Z_3 Z_4 Z_5 Z_6 Z_7$ and $X_L = X_1 X_2 X_3 X_4  X_5 X_6  X_7$, share an odd number (seven) of qubits, and thus fulfill the correct anti-commutation relation $\{ X_L, Z_L\} = 0$. Furthermore, as required, they commute with the six stabilizers (or generators) $S_x^{(i)}$ and $S_z^{(i)}$, $i=$1, 2, 3, of the code, as the logical operators share 4 physical qubits with each stabilizer operators. Thus, logical quantum states $\ket{\psi}_L$ remain within the code space under these logical operations. The logical $Y_L$ operation is given by the sequential application of both an $X_L$ and a $Z_L$ operation, $Y_L = + i X_L Z_L = - Y_1 Y_2 Y_3 Y_4 Y_5 Y_6  Y_7$. The logical $Z_L$-gate can be implemented in our setup (see Sec.~\ref{sec:SI_coherent_gates}) by single-ion $Z$ rotations, $Z_L = \prod_{i=1}^7 U_Z^{(i)}(\pi)$. The logical $X_L$ is realized by a collective local rotation around the $X$-axis, $X_L = U(\pi, 0)$.

We realize the logical Hadamard gate $H_L = H_1H_2 H_3 H_4 H_5 H_6 H_7$ by a collective local $Y$ rotation, followed by single-ion $Z$-rotations: $H_L = \prod_{i=1}^7 U_Z^{(i)}(\pi) U(-\pi/2, \pi/2)$. Note that, in principle, for some input states and at the end of sequences of encoded Clifford gates, directly before the fluorescence measurement in the computational ($Z$) basis, the application of the laser pulses to physically realize the $Z$ rotations on all ions could be omitted, as they would (ideally) not change the logical state and measurement outcomes. However, during the time it takes to apply the pulses to the ions, the encoded logical qubit will be exposed to decoherence, and furthermore the addressed pulses are associated with small addressing errors due to cross-talk with neighboring ions (see Sec.~\ref{sec:SI_decoupling}). Thus, in order to properly take into account these effects and to study in an un-biased way without additional assumptions the effect of imperfections in the Clifford operations as elementary building blocks, we always apply the full sequence of listed laser pulses for all cases where a logical $H_L$ gate operation is executed. We furthermore emphasize that -- in contrast to other quantum codes, such as the 5-qubit code \cite{s-nielsen-book}, where the logical Hadamard operation requires a 5-qubit entangling unitary \cite{s-zhang-prl-109-100503} -- here, owing to the transversal character of the logical Hadamard gate operation in the 7-qubit code, the implementation is achieved using exclusively non-entangling local operations.

Finally, we implement the logical phase or $K_L$ gate in a transversal way. As compared to other topological codes, it is a distinguishing feature of 2D color codes (embedded in 4-8-8 lattices, see discussion in Sec.~\ref{sec:SI_topological_order} and Fig.~\ref{fig:lattices} for details), that the latter codes enable a transversal implementation of the $K_L$ gate operation \cite{s-bombin-prl-97-180501}, not requiring the technique of magic-state injection via an ancillary qubit \cite{s-bravyi-pra-71-022316} nor multi-qubit entangling operations such as, e.g.~in the non-transversal 5-qubit code \cite{s-nielsen-book}. Note that the logical $K_L$ gate operation is required to fulfill $K_L X_L K_L^\dagger = Y_L = i X_L Z_L$. Note that for $K_L = K_1K_2 K_3 K_4 K_5 K_6 K_7$, however, one obtains $K_L X_L K_L^\dagger = - i X_L Z_L$; thus $K_L$ defined in this way acts as $K_L^\dagger$ within the code space. This detail can be readily cured by the redefinition $K_L := \prod_{i=1}^7 K_i^\dagger$. We implement the $K_L$ gate operation by bit-wise $Z$-rotations, $K_L = \prod_{i=1}^7 U_Z^{(i)}(-\pi/2)$.

\subsection{Preparation of the six eigenstates of the logical operators $X_L$, $Y_L$ and $Z_L$}
According to the quantum circuits shown in Fig.~3A and B of the main text, we use up to three Clifford gate operations to prepare the six logical states lying along the axes of the logical Bloch sphere, starting with the logical qubit initially in state $\ket{0}_L$. The experimentally generated states, required logical gate operations, average stabilizer expectation values ($\langle S_i \rangle = \frac{1}{6}\sum_{i=1}^3 (\langle S_x^{(i)} \rangle + \langle S_z^{(i)} \rangle)$), and length $L$ of the logical Bloch vector ($L = \sqrt{\langle X_L \rangle^2 + \langle Y_L \rangle^2 + \langle Z_L \rangle^2}$) for each state are:
\begin{itemize}
\item
$\ket{0}_L$: $\langle S_i \rangle$ = 0.42(1), $L$ = 0.38(6) -- initial encoded state, preparation requires no Clifford gate operation.
\item
$\ket{1}_L$: $\langle S_i \rangle$ = 0.54(1), $L$ = 0.57(6) -- requiring one $X_L$ gate operation.
\item
$\ket{-_x}_L$: $\langle S_i \rangle$ = 0.49(2), $L$ = 0.58(2) -- requiring one $X_L$ and one $H_L$ gate operation.
\item
$\ket{+_x}_L$: $\langle S_i \rangle$ = 0.52(1), $L$ = 0.48(2) -- requiring one $H_L$ gate operation.
\item
$\ket{+_y}_L$: $\langle S_i \rangle$ = 0.39(1), $L$ = 0.42(2) -- requiring one $H_L$ and one $K_L$ gate operation.
\item
$\ket{-_y}_L$: $\langle S_i \rangle$ = 0.42(1), $L$ = 0.38(2) -- requiring one $H_L$, one $K_L$, and one $X_L$ gate operation.
\end{itemize}

\subsection{Longer sequences of encoded gate operations and decay of coherences of the logical qubit}
After preparing the encoded qubit in the logical state $\ket{-_y}_L$ by three Clifford gate operations (see previous paragraph and sequence shown in Fig.~3B of the main text), we applied up to 10 additional logical $X_L$ gate operations to induce flips of the logical qubit between the +1 and -1 eigenstates of $Y_L$. A weighted exponential fit of the form
$A \exp{(-n_{gate}/B)}$, with $n_{gate}$ the number of logical gates, into the $\langle Y_{L} \rangle$ expectation values yields a decay rate of the average expectation value of 3.8(5)\% per gate.
This decay is consistent with what we expect from the accuracy with which collective resonant $\pi$-rotations $U(\pi,0)$ can be implemented on a string of 7 ions in our setup.
A fidelity as high as about 99.6\% of a single collective resonant $\pi$-rotation per ion would already lead to a fidelity loss of $\approx 3.8\%$ per gate operation.
Here, two error sources dominate the measured fidelity loss per gate: (i) The relative intensity inhomogeneity of the global laser beam across the ion string $(\approx 1\%)$, lowering the Rabi-frequencies at the ions located at the edge of the string; and (ii) thermal occupation of higher motional modes, which leads to decoherence of the global Rabi oscillations~\cite{s-schindler-njp-15-123012}. The application of the pulses for 10 $X_L$ gate operations is realized in a duration of 200$\mu$s. This is a factor of 18 shorter than the $1/e$ time of $3.6(6)$\,ms on which logical coherences, as indicated by the expectation value $\langle X_L \rangle$, of the qubit initially prepared in $\ket{+_x}_L$, decay. The corresponding decay constant has been obtained by a weighted exponential fit into the $\langle X_L \rangle$ stabilizer expectation values. Thus, for the executed circuit of encoded quantum gates, imperfections in the logical Clifford gate operations dominate over the effect of the bare decoherence of the logical qubit.
A more extensive study of the decoherence properties of the logical qubit, as well as an exhaustive characterization of the set of encoded Clifford gate operations, e.g.~by means of randomized benchmarking techniques \textit{for encoded logical gate operations} or by other techniques for quantum process characterization, lies beyond the scope of the present work and will be the focus of future research.

\end{document}